\PassOptionsToPackage{pdfpagelabels=false}{hyperref} 
\documentclass[fleqn,usenatbib,useAMS]{mnras}
\usepackage{xspace}

\usepackage{hyperref}

\usepackage{graphicx}
\usepackage{inputenc}
\usepackage[Symbol]{upgreek}
\usepackage[T1]{fontenc}
\usepackage{lmodern}
\usepackage{inputenc}
\usepackage{epsfig}
\usepackage{hyperref}
\usepackage{amsmath}
\usepackage{lmodern}
\usepackage[T1]{fontenc}
\usepackage{natbib}
\usepackage{relsize}
\usepackage{ae,aecompl}
\usepackage{graphicx}	
\usepackage{amsmath}	
\usepackage{amssymb}	
\usepackage{times}

\newcommand{\kms}{km~s$^{-1}$ }
\newcommand{\kmsp}{km~s$^{ -1}$}
\newcommand{\oi}{\ion{O}{i}~}

\newcommand{\siii}{\ion{Si}{ii}~}
\newcommand{\siiii}{\ion{Si}{iii}~}
\newcommand{\sii}{\ion{S}{ii}~}

\newcommand{\siiv}{\ion{Si}{iv}~}

\newcommand{\oip}{\ion{O}{i}}

\newcommand{\siiip}{\ion{Si}{ii}}

\newcommand{\siiiip}{\ion{Si}{iii}}

\newcommand{\siip}{\ion{S}{ii}}

\newcommand{\siivp}{\ion{Si}{iv}}

\newcommand{\wise}{{\it WISE} }

\newcommand{\galex}{{\it GALEX} }
\newcommand{\galexp}{{\it GALEX}}

\newcommand{\mout}{$\dot{M}_{\text{o}}$ }
\newcommand{\mstar}{$M_\ast$ }
\newcommand{\mstarp}{$M_\ast$}

\newcommand{\moutp}{$\dot{M}_{\text{o}}$}

\newcommand{\vcircp}{$v_{\text{circ}}$}

\newcommand{\sfr}{M$_\odot$~yr$^{-1}$ }
\newcommand{\sfrp}{M$_\odot$~yr$^{-1}$}

\newcommand{\zo}{Z$_\mathrm{o}$ }
\newcommand{\zop}{Z$_\mathrm{o}$}
\newcommand{\zism}{Z$_\mathrm{ISM}$}

\newcommand{\no}{n$_{\mathrm{H},0}$}
\newcommand{\mzp}{$\dot{\mathrm{M}}_\mathrm{Z}$}
\newcommand{\mz}{$\dot{\mathrm{M}}_\mathrm{Z}$ }

\newcommand{\hi}{\ion{H}{i}}
\newcommand{\lya}{Ly$\alpha$}
\newcommand{\zs}{Z$_\text{s}$}
\newcommand{\rft}{R$_{42}$}
\newcommand{\rtt}{R$_{32}$}
\newcommand{\sfrc}{SFR$_\mathrm{COS}$}
\newcommand{\ebv}{E$_\mathrm{S}$(B-V)}

\newcommand{\nhi}{$N_\mathrm{HI}$}
\defcitealias{chisholm15}{Paper I}	
\defcitealias{chisholm16}{Paper II}
\defcitealias{chisholm16b}{Paper {\small III}}	
\defcitealias{chisholm17}{Paper {\small IV}}	
\newcommand{\megasaura}{M\textsc{eg}a\textsc{S}a\textsc{ura}}

\begin{document}

\title[Outflows shape the mass-metallicity relation]{Metal-enriched galactic outflows shape the mass-metallicity relationship}
\author[Chisholm et al.]{J. Chisholm$^{1}$\thanks{Contact email: John.Chisholm@unige.ch}, C. Tremonti$^{2}$, and C. Leitherer $^{3}$\\
$^{1}$  Observatoire de Gen\`{e}ve, Universit\`{e} de Gen\`{e}ve, 51 Ch. des Maillettes, 1290 Versoix, Switzerland \\
$^{2}$Astronomy Department, University of Wisconsin, Madison, 475
  N. Charter St., WI 53711, USA \\
$^{3}$Space Telescope Science Institute, 3700 San Martin Drive, Baltimore, MD 21218, USA \\
}
\pubyear{2018}
\label{firstpage}
\pagerange{\pageref{firstpage}--\pageref{lastpage}}
\maketitle
\begin{abstract}
The gas-phase metallicity of low-mass galaxies increases with increasing stellar mass ($M_\ast$) and is nearly constant for high-mass galaxies. Theory suggests that this tight mass-metallicity relationship is shaped by galactic outflows removing metal-enriched gas from galaxies. Here, we observationally model the outflow metallicities  {of the warm outflowing phase from} a sample of seven local star-forming galaxies  {with stellar masses between 10$^{7}$--10$^{11}$~M$_\odot$}. We estimate the outflow metallicities using four weak rest-frame ultraviolet absorption lines, the observed stellar continua, and photoionization models. The outflow metallicity is flat with $M_\ast$, with a median metallicity of $1.0\pm0.6$~Z$_\odot$. The observed outflows are metal-enriched: low and high-mass galaxies have outflow metallicities 10-50 and 2.6 times larger than their ISM metallicities, respectively.  {The observed outflows are mainly composed of entrained ISM gas {with} at most 22\% of the metals directly coming from recent supernovae enrichment}. The metal outflow rate shallowly increases with $M_\ast$, as $M_\ast^{0.2 \pm 0.1}$, because the mass outflow rate shallow increases with $M_\ast$.  Finally, we normalize the metal outflow rate by the rate at which star formation retains metals to calculate the metal-loading factor. The metal-loading factor inversely scales with $M_\ast$. The normalization and scaling of the metal-loading factor agree with analytic expressions that reproduce observed mass-metallicity relations. Galactic outflows fundamentally shape the observed mass-metallicity relationship.
\end{abstract}
\begin{keywords}
ISM: jets and outflows, galaxies: evolution, galaxies: formation, ultraviolet: ISM
\end{keywords}

\section{INTRODUCTION}

Many galaxy properties are highly correlated. The star formation rate (SFR) sublinearly increases with the stellar mass (\mstarp), at nearly all redshifts, as the star formation main sequence \citep{brinchmann2004, salim07, noeske07, elbaz07}; the stellar mass increases with the halo mass \citep{moster10}; the gas-phase metallicity scales with \mstar \citep{Lequeux, skillman, tremonti04, erb06, berg, andrews13, zahid}; and the gas-phase metallicity correlates strongly with \mstar {\it and} SFR, as the so-called fundamental metallicity relation \citep{mannucci, lopez10}. By connecting different galactic components, these relations suggest that physical processes shape the formation and evolution of galaxies. 
\\

Among these relations, the stellar mass gas-phase metallicity relation (MZR) has perhaps received the most attention. Observations suggest that the gas-phase metallicity (typically measured from strong optical oxygen nebular emission lines as 12+log(O/H)) increases with the \mstar of low-mass galaxies, but flattens to a nearly constant value at large \mstar \citep{tremonti04}. While this qualitative shape is nearly universally observed, the normalization (i.e. the absolute metallicity values and the \mstar where the relations flatten) strongly depends on how the metallicity is calculated \citep{kewley08}. Absolute metallicity calibrations can vary up to 0.7~dex for different calibration systems. 
\\

The MZR connects host galaxy observable properties to fundamental processes in galaxy evolution. To start, the quantities relate the integrated conversion of gas into stars (the total \mstarp) to the integrated amount of gas processed through, and therefore enriched by, stars (the metallicity). While both \mstar and metallicity are by-products of star formation, they represent different phases of baryonic matter.

The simplest attempt to theoretically explain the MZR is the closed box scenario where a galaxy initial consists of metal-poor gas which gravitationally collapses to form stars \citep[see the review in ][]{tinsley80}. No gas enters or leaves the galaxy in this scenario. In this simplistic situation, high-mass stars produce metals through nucleosynthesis, returning those metals, at a yield of $y$, to the surrounding gas through stellar winds and supernovae. With these assumptions, the gas-phase metallicity (\zism) increases as gas is processed through stars as
\begin{equation}
Z_\mathrm{ISM} = y \ln\frac{M_g + M_\ast}{M_g} = y\ln\mu^{-1} ,
\label{eq:closed}
\end{equation} 
where M$_g$ is the gas mass and $\mu$ is the gas mass fraction \citep{searle}. While simple, \autoref{eq:closed} catastrophically fails to reproduce the metal content of stars in the solar neighborhood \citep[the G-dwarf problem;][]{schmidt, vandenbergh, pagel75}; other physics must shape the metallicities of galaxies.

Galaxies do not evolve as closed systems. Relatively metal poor gas accretes onto galaxies from the intergalactic medium (IGM), diluting the metallicity  \citep{larson72}. Accretion rates larger than the SFR may decrease the metallicity of a galaxy to levels observed in low-mass galaxies \citep{edmunds, koppen, garnett}. However, these excessive accretion events lead to bursts of star formation that enrich the galaxy back to the closed box value (\autoref{eq:closed}).  Accretion cannot decrease metallicities over an extended duration \citep{dalcanton}. 

While galaxies acquire gas and metals from the IGM, they also lose metals through galactic outflows \citep{heckman90, heckman2000, veilleux05}. Supernovae, high-energy photons, cosmic rays, and stellar winds inject energy and momentum into the ISM, accelerating gas out of galaxies. Since low-mass galaxies have shallower gravitational potentials, star formation powered galactic outflows more efficiently remove metals from low-mass galaxies to produce the MZR \citep{larson74, dekel86, tremonti04, dalcanton, finlator08, peeples11, lilly13}. In fact, analytic work by \citet{dalcanton} showed that galactic outflows are the most likely physical process to decrease the metallicites of low-mass galaxies. To produce the MZR, galactic outflows must remove more metals than their star formation retains \citep{dalcanton, peeples11}. By preferentially driving metals out of galaxies, stellar feedback may decrease the gas-phase metallicity of low-mass galaxies and shape the MZR. 

Galactic outflows are ubiquitously observed in star-forming galaxies at all epochs as broad blue shifted emission lines \citep{lynds, sharp, arribas2014} and interstellar metal absorption lines \citep{heckman90, heckman2000, pettini2002, shapley03, rupke2005b, martin2005, weiner, chen10, martin12, rubin13, bordoloi, heckman15}. However, it is challenging to measure the amount of metals and mass removed by galactic outflows (the mass outflow rate; \moutp) because outflows have uncertain geometries, ionization corrections, and metallicities; all of which may add an order of magnitude uncertainty to \mout estimates \citep{murray07, chisholm16b}. Observations need to determine if the metal outflow rate (\mzp) of galactic outflows is sufficient to shape the MZR.

In a series of papers, we have explored galactic outflows from local star-forming galaxies using restframe ultraviolet (UV) down-the-barrel spectroscopy from the Cosmic Origins Spectrograph (COS) on the {\it Hubble} Space Telescope. We started this series by characterizing the host galaxies, the stellar continua, and the galactic outflow properties using a single \siii absorption line \citep[][ hereafter Paper I]{chisholm15} and multiple absorption lines spanning a range of ionization states \citep[][ hereafter Paper II]{chisholm16}. In \citet{chisholm16b} (hereafter Paper III) we used the observed column densities to map out the ionization structure, outflow metallicity, and \mout of a single galaxy, NGC~6090. \citet{chisholm17} (hereafter paper IV) extended the \mout calculation to a sample of 7 galaxies, with spectra that have sufficient signal-to-noise ratios to explore how \mout scales with \mstarp. 

\begin{table*}
\caption{{Derived galaxy properties. Column 1 is the galaxy name; column 2 is the stellar mass (\mstarp); column 3 is the galactic circular velocity (\vcircp) calculated using a Tully-Fischer relation \citep{reyes}; column 4 is the star formation rate of the entire galaxy (SFR); column 5 is the SFR within the COS aperture (SFR$_\mathrm{COS}$); column 6 is the oxygen abundance (12+log(O/H)); column 7 is the stellar continuum metallicity (Z$_\mathrm{s}$); column 8 is the light-weighted stellar age; column 9 is the stellar continuum attenuation (E$_\mathrm{S}$(B-V)) corrected for foreground reddening \citep{schlegel}; column 10  {gives the distances used and, when appropriate, the references for the distances}; column 11 lists the \textit{HST} proposal ID for each observation; and in column 12 we give references to other papers that use the COS data. Columns 7-9 were determined from the stellar continuum fits in \autoref{cont}.Metallicity references: a) \citet{cortijo}, b) \citet{izotov97}, c) \citet{izotov98} d) \citet{james13},  e) \citet{ostlin}, f) \citet{perez}.}  {Distance references in column 10 are: ($\alpha$) \citet{aloisi05} and ($\beta$) \citet{fiorentino}; the rest are computed from the redshift and the Hubble flow.} References in column 12 are: (1) \citet{alexandroff}, (2) \citet{duval}, (3) \citet{france2010}, (4) \citet{fox2013}, (5) \citet{fox14}, (6) \citet{hayes14}, (7) \citet{heckman15}, (8) \citet{james}, (9) \citet{claus2012}, (10) \citet{ostlin}, (11) \citet{pardy}, (12) \citet{Rivera}, (13) \citet{richter2013}, and (14) \citet{wofford2013}.}
\resizebox{\textwidth}{!}{
\begin{tabular}{lccccccccccc}
\hline
Galaxy name & log(\mstarp) & \vcircp   & SFR & SFR$_\mathrm{COS}$ & 12+log(O/H) & Z$_\mathrm{s}$ & <Age> & E$_\mathrm{S}$(B-V) & D & Proposal ID & References  \\
 & [log(M$_\odot$)]  & [\kmsp] & [M$_\odot$~yr$^{-1}$]& [M$_\odot$~yr$^{-1}$] & [dex] & [Z$_\odot$] &[Myr] & [mag]& [Mpc] &  &  \\
 (1)  & (2) & (3)& (4)& (5)& (6) & (7)  &(8) & (9) & (10) & (11) & (12)  \\ 
\hline
SBS~1415+437 &6.9 & 18& 0.016 & 0.016 & 7.59$^c$ & 0.05 & 3.7 & 0.10 &13.6$^{\alpha}$& 11579  &8 \\
1~Zw~18 &7.2 & 21 &  0.023 & 0.023 & 7.22$^b$ & 0.05 & 7.3 & 0.09 &18.2$^{\beta}$ & 11579 &  8 \\
MRK~1486 &9.3 & 82 &3.6 & 2.5&  7.80$^e$ & 0.2 &8.0 & 0.22 & 148& 12583 &  2, 6, 10, 11, 12 \\
KISSR~1578 & 9.5 & 94 & 3.7 & 2.1 & 8.07$^e$ & 0.2 & 5.5 & 0.13 & 122& 11522 & 3, 14 \\
Haro~11 & 10.1 & 137 & 26 & 12 & 7.80$^d$ & 0.4 & 10.1 &0.13 & 89&13017& 1, 7   \\
NGC~7714 &10.3 & 156 & 9.2 & 3.1 &  8.23 $^f$ &1 & 4.3 & 0.31 & 40& 12604 & 4, 5, 13\\
NGC~6090 & 10.7 & 202& 25& 5.5 & 8.40$^a$& 1 & 4.5 & 0.32& 128&  12173 &9\\
\end{tabular}
}
\label{tab:sample}
\end{table*}

Here, we conclude the series of papers by examining, for the first time, the outflow metallicities and {\it metal} outflow rates of star-forming galaxies. We begin by reviewing the host galaxy properties and the COS observations of our sample (\autoref{data}). We then summarize the steps to determine the metal outflow rates of these seven galaxies, including: the measurement and removal of the stellar continuum (\autoref{cont}), the profile fitting and mass outflow rate calculation (\autoref{proffit}), and the ionization modeling (\autoref{cloudy}). Trends with our metal outflow rates are subsequently explored (\autoref{results}). We then compare our estimates to previous observations (\autoref{prev}), discuss the role outflows play in shaping the observed MZR (\autoref{mzrel}), and  explore some limitations and future steps of this study (\autoref{further}). In \autoref{cloudy_obs}, we discuss the full parameter space of our photoionization models. These observations provide the first empirical evidence that galactic outflows shape the mass-metallicity relationship. 
\\\

Throughout this paper we use a solar gas-phase metallicity of 0.0142 and 12+log(O/H)$_\odot$ of 8.69 \citep{asplund}. Stellar solar metallicity is 0.02 \citep{claus99}.  {Additionally, we use the Chabrier initial mass function \citep[IMF;][]{chabrier} when determining host galaxy properties and a Kroupa IMF \citep{kroupa} when fitting the stellar continua in \autoref{cont}.}

\section{DATA}
\label{data}
\subsection{Sample and host galaxy properties}
\label{sample}

We drew the sample of galaxies from the (HST) archive of ultraviolet (UV) spectra observed with COS \citep[][]{cos}. Following \citetalias{chisholm17}, we took the seven star-forming galaxies from the COS archive that have a signal-to-noise ratio greater than 5 near the \siivp~1402\AA\ doublet, do not have geocoronal emission or Milky Way absorption near the \siiip~1304\AA\ line, and are well-fit by the model detailed in \autoref{proffit}. \oip~1306\AA\ geocoronal emission contaminates the \siiip~1304\AA\ line for many of the low redshift galaxies within the COS archive, making it challenging to constrain the ionization structure without this weak (the least saturated \siii line in the wavelength regime) \siii line (see \autoref{cloudy}). Previous references and the original \textit{HST} proposal IDs for each galaxy are listed in \autoref{tab:sample}.

These spectra have moderate spectral resolution, with full width at half-maximums between 21-58~\kms as measured from the Milky Way absorption lines. We observe both strong and weak interstellar medium absorption lines, which diagnose the conditions and velocities of galactic outflows. These galaxies were selected such that their \oip~1302\AA, \siiip~1304\AA, and \siivp~1402~\AA\ absorption lines have median velocities blueward of 0~\kmsp, relative to the stellar continuum features (see \autoref{cont}), at the $1\sigma$ level (see \autoref{fig:s42}). These kinematics ensure that the COS observations probe galactic outflows and not the static interstellar medium (but see the discussion in \autoref{prev}). 

\begin{figure}
\includegraphics[width = .5\textwidth]{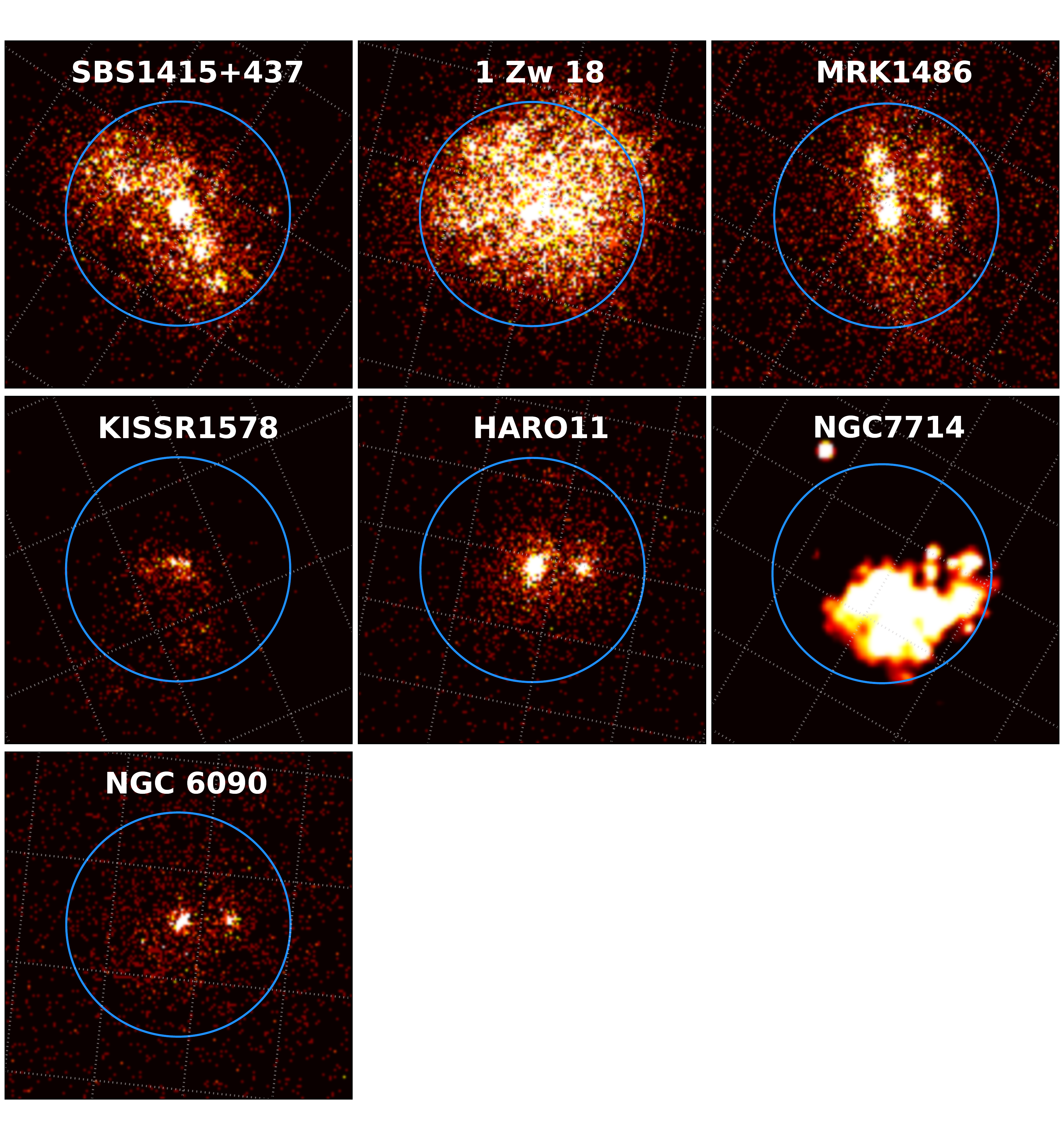}
\caption{ {COS acquisition images (except for NGC~7714 where we show an HST/WFC3 F300X image with rest wavelength of 2800~\AA; HST proposal ID: 12170, PI:  Gal-Yam) for the seven galaxies included in the sample, as denoted by the name in each panel. The COS 2\farcs5 diameter aperture is shown as a blue circle. MRK~1486, Haro~11, and NGC~6090 were observed with Mirror~B, creating the, unphysical, ghost duplicate image seen in these panels.}}
\label{fig:acq_image}
\end{figure}

The sample represents the highest quality restframe UV observations in the local universe, and spans nearly the entire local host galaxy parameter range.  {We calculated the stellar mass (\mstarp) using ancillary {\it WISE} restframe 3.6$\mu$m luminosities \citep{wise}, assuming the calibration from \citet{querejeta} which includes a [3.6$\mu$m]-[4.5$\mu$m] color mass-to-light ratio. The luminosities were calculated using the distances in column 10 of \autoref{tab:sample}. The two lowest mass galaxies, 1~Zw~18 and SBS~1415+437, are not in the Hubble flow and we used distances calculated from the tip of the red giant branch and Cepheid pulsation periods \citep{aloisi05, fiorentino}. The rest of the sample is in the Hubble flow and we calculated the luminosity distance using the SDSS redshift \citepalias{chisholm15}. } The star formation rates (SFR) were calculated using a combination of {\it GALEX} UV photometry \citep{galex} and {\it WISE} 22$\mu$m observations using the  {scaling} relation from \citet{buat11}:
\begin{equation}
    \text{SFR} = \text{SFR}_\text{UV} + 0.83 \text{SFR}_\text{IR},
\end{equation}
 {where SFR$_\text{UV}$ is calculated using the scaling relation from the \galex\ UV flux \citep{kennicutt2012} and SFR$_\text{IR}$ is calculated from the \wise 22~$\mu$m luminosity \citep{jarrett2011}. This relationship includes the 22$\mu$m observations, corresponding to dust heated by obscured massive stars, as well as the direct FUV continua of unobscured massive stars. { Accounting for both obscured and unobscured star formation}} is important because the sample includes low metallicity star-forming galaxies and massive dusty mergers (see \autoref{tab:sample}).  {Both the \mstar and SFR calibrations assumed a Chabrier IMF \citep{chabrier}}. As described in \citet{jarrett2013}, we assumed that the SFRs have a 20\% error, but SFRs, especially from bursty star formation histories, are more uncertain \citep[for a discussion about 1~Zw~18 see][]{annibali}.

COS is a 2\farcs5 diameter circular aperture spectrograph, and many of these COS observations only probe a fraction of the total star formation within the galaxy  {(see the patchy nature of the UV emission in the COS NUV acquisition images in \autoref{fig:acq_image})}. To correct for this aperture effect, we calculated the SFR within the COS aperture (\sfrc) as the total SFR (\wise plus \galexp) times the flux ratio of the COS and the \galex observations (F$_\mathrm{COS}$/F$_\mathrm{GALEX}$). This accounts for aperture effects, and produces a SFR within the COS aperture. We use \sfrc\ while studying the outflow properties because recent studies suggest that outflow properties, like velocity, vary with the local star formation and not the global star formation \citep[][although see \citet{james18}]{bordoloi16}. All host galaxy properties are tabulated in \autoref{tab:sample}. 

The interstellar medium metallicities (\zism) were taken from the literature (see the references in \autoref{tab:sample}). We were careful to use metallicities as close as possible to the location of the COS aperture with as homogeneous of calibration procedures as possible: six of the seven galaxies have T$_e$ based 12+log(O/H) calibrations. The only galaxy without a T$_e$ based metallicity, NGC~6090, is calculated using the strong-line O3N2 calibration from \citet{marino} which is empirically tied to the T$_e$ scale. While we attempted to use homogeneous metallicity calibrations, metallicity calibrations still have large relative differences \citep{kewley08}. Thus, we assumed 12+log(O/H) errors of 0.15~dex instead of the reported errors. This error is the relative RMS that \citet{kewley08} found between different metallicity calibrations, and accounts for the differences in the metallicity calibrations. The individual parameters are highly correlated within the small sample (O/H~$\propto \mathrm{Z}_\mathrm{ISM} \propto M_\ast^{0.32}$ and SFR~$ \propto {M}_\ast^{0.75}$), as expected from the MZR and the fundamental metallicity relation \citep{mannucci, lopez10}. 

\subsection{Spectral reduction}
\label{reduction}
The COS spectra were downloaded from MAST and processed through {\small CalCOS} version 2.20.1. The individual exposures were then combined using the methods outlined in \citet{wakker2015}, where the wavelength calibration for each galaxy was defined using strong Milky Way absorption lines and Milky Way 21~cm emission from the LAB survey \citep{kalberla}. The flux errors were calculated using the raw counts that were converted into flux units using the reported gain in \citet{coshandbook}. We binned the spectra by 5 pixels (a velocity spacing of 10~\kmsp), and smoothed the spectra by 3 pixels to produce the final reduced spectra. The median signal-to-noise ratio of the sample is 11 per resolution element near the \siivp~1402\AA\ doublet. 

\section{SPECTRAL ANALYSIS AND OUTFLOW METALLICITY CALCULATION}
\label{calc}
We now use the reduced COS spectra to determine the outflow metallicities. First, we fit and remove the stellar continuum from the down-the-barrel COS observations using {\small STARBURST99} stellar continuum model fits (\autoref{cont}). Once the stellar continuum features are removed from the spectra, we use the \siivp~1402\AA\ interstellar absorption doublet to determine how the covering fraction and density evolves with velocity (\autoref{proffit}). These are important input parameters for the ionization modeling and also describe the mass outflow rates. Finally, in \autoref{cloudy} we describe the photoionization models that estimate the outflow metallicities. These photoionization models use the observed stellar continua within the COS aperture as the ionizing source and the density distribution from the \siiv interstellar absorption lines. These methods were first outlined in \citetalias{chisholm16b}, and we refer the reader to that paper for a detailed description of their derivation.  

\subsection{Stellar continuum fitting}
\label{cont}
\begin{figure*}
\includegraphics[width = \textwidth]{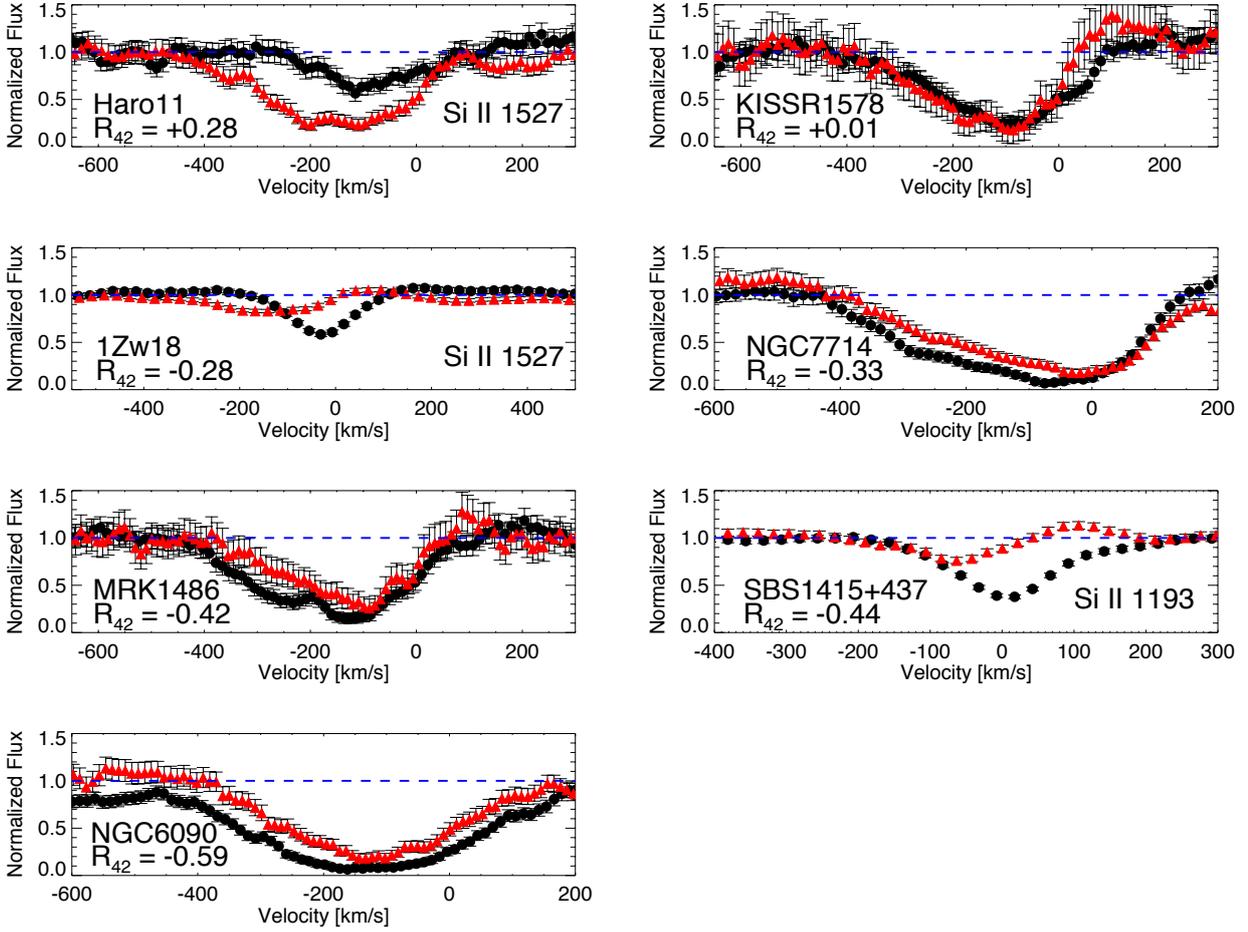}
\caption{Plots of the {\small STARBURST99} stellar continuum normalized \siiip~1190\AA\ (black circles) and \siivp~1402\AA\ (red triangles) absorption profiles. The names of each galaxy are in the lower left of each panel. Zero-velocity corresponds to the restframe of the background stellar continuum (see \autoref{cont}). The \siiip~1190\AA\ and \siivp~1402\AA\ lines have similar $f$-values (0.277 and 0.255, respectively) and the difference in column densities of the two transitions is the difference in integrated absorption depth, if both transitions have the same covering fraction. The galaxies are ordered in descending order of the \siivp-to-\siii column density ratio (\rft$=\log(N_\mathrm{\siivp}/N_\mathrm{\siiip}$)). Each \rft\ value is included below the name of the galaxy. Galaxies with positive \rft\ have larger \siiv column densities than \siii column densities (e.g. Haro~11). Haro~11 and 1~Zw~18 have geocoronal emission and Milky Way absorption at 1190\AA. Instead, we use the 1527\AA\ line, which has an $f$-value half the size of \siivp~1402\AA. Similarly, the Milky Way \siiip~1193\AA\ line contaminates the \siiip~1190\AA\ line of SBS~1415+437. Instead, we plot the \siiip~1193\AA\ line, which has an $f$-value twice as large as \siivp~1402\AA. The galaxies that do not use \siiip~1190\AA\ have the associated \siii transitions labeled in the lower right.}
\label{fig:s42}
\end{figure*}
The COS observations were made of individual UV-bright star clusters, such that the outflow absorption lines are imprinted on top of the background stellar continuum. The shape of the stellar continuum drastically affects the continuum level of interstellar absorption lines such as \siiv (see fig. 2 in \citetalias{chisholm16}). Therefore, we fit the background stellar continuum using a linear combination of multiple-age, fully theoretical {\small STARBURST99} stellar continuum models created using the WM-Basic method \citep{claus99, claus2010} and the Geneva stellar evolution tracks with high-mass loss \citep{geneva94}. We used a Kroupa IMF with a high-mass cutoff at 100~M$_\odot$ and a power-law exponent of 1.3 (2.3) for the low- (high-) mass end. 

The far-ultraviolet stellar continuum is dominated by stars younger than 10~Myr; the shape and features of the stellar continua in the wavelength regime we used, 1200--1450\AA, do not change appreciably for  stellar populations older than 40~Myr \citep{demello}. Further, the number of ionizing photons -- an important input for our ionization models -- decreases with increasing stellar age. Therefore, we used a linear combination of 10 single-age fully theoretical {\small STARBURST99} stellar continuum models, with ages between 1--40~Myr, to describe the background stellar continuum as multiple bursts of star formation \citepalias{chisholm15}. The stellar continuum metallicity (\zs) strongly alters the number of ionizing photons emitted by a stellar population because metals in stellar atmospheres absorb and scatter ionizing photons \citepalias[eq. 12 in][]{chisholm16}. We allowed for the 5 {\small STARBURST99} stellar continuum metallicities (0.05, 0.2, 0.4, 1, and 2~Z$_\odot$), and determined a best fit \zs\  using a $\chi^2$ test.

While the stellar continuum describes most of the observed continuum shape, dust along the line-of-sight changes the spectral shape by absorbing and scattering continuum photons to longer wavelengths. Consequently, we accounted for dust attenuation by reddening the {\small STARBURST99} models with a Calzetti reddening law \citep{calzetti} with a varying stellar continuum attenuation parameter (E$_\mathrm{S}$(B-V)). The final \ebv\ measurements were corrected for foreground extinction using the maps from \citet{schlegel}. 

We simultaneously fit for E$_\mathrm{S}$(B-V) and the linear combination of the 10 single-age {\small STARBURST99} models using the non-linear least squares fitter {\small MPFIT} \citep{mpfit}. Finally, we cross-correlated the {\small STARBURST99} fit with the observed data to set the zero-velocity (systemic) of the background stellar population.

Either the stellar continuum fits (fig.~2 in \citetalias{chisholm15} and \citetalias{chisholm16}) or the stellar continuum normalized spectra (fig.~3 in \citetalias{chisholm15} or fig.~2 in \citetalias{chisholm16b}) are shown in previous publications. We show the stellar continuum normalized \siii and \siiv absorption profiles in \autoref{fig:s42}. We have shown other line profiles from these galaxies in previous papers (fig~.4 and fig.~6 in \citetalias{chisholm15}, fig.~3 in \citetalias{chisholm16}, and fig~2. in \citetalias{chisholm16b}). The fitted mean light-weighted stellar age, E$_\mathrm{S}$(B-V), and \zs\ are given in \autoref{tab:sample}. These parameters are inputs for the photoionization models in \autoref{cloudy}. 

\subsection{Profile fitting and mass outflow rate calculations}
\label{proffit}

Both the ionizing source and density distribution influences the ionization structure. In \autoref{cont}, we fit the stellar continuum to describe the ionization source of the outflows. Here, we describe how we use the observed \siiv line profiles to characterize the density distribution (n$\left(v\right)$) of the outflow.

In \autoref{fig:s42} the \siiv and \siii line profiles show a variety of shapes and widths. Only the \siii from NGC~7714 and NGC~6090 appear black at any velocity; the other galaxies and transitions are either optically thin or have a covering fraction less than unity. Consequently, both optical depth ($\tau$) and covering fraction ($C_f$) shape the outflow line profiles. The degeneracy between these two parameters is characterized by the radiative transfer equation, which describes the observed velocity-dependent  flux ($F(v)$) as:
\begin{equation}
F(v) = 1- C_f\left(v\right) + C_f\left(v\right) e^{-\tau\left(v\right)} .
\label{eq:radiative}
\end{equation}
The degeneracy between $\tau(v)$ and $C_f(v)$ is broken using a doublet because \autoref{eq:radiative} becomes a system of two equations with two unknowns since both lines have the same $C_f(v)$ and $\tau(v)$ is tied by the ratio of the oscillator strengths of the two transitions. Using the \siivp~1402\AA\ doublet, we solved for the observed $C_f(v)$ and $\tau(v)$ using \autoref{eq:radiative} (see the Appendix of \citetalias{chisholm17} for these distributions).

We used simple, physically motivated, models to derive the variation of $C_f(v)$ and $\tau(v)$. First, we assumed that $C_f(v)$ varies as a power-law with radius ($r$) such that:
\begin{equation}
C_f (r) = C_f(\mathrm{R}_\mathrm{i}) \left(\frac{r}{\mathrm{R}_\mathrm{i}}\right)^{\gamma} = C_f (\mathrm{R}_\mathrm{i}) x^\gamma ,
\label{eq:cf}
\end{equation}
where $C_f(R_\mathrm{i})$ is the $C_f$ at the initial radius ($\mathrm{R}_\mathrm{i}$) and $\gamma$ is the exponent that describes how $C_f$ changes with $r$. We simplified this relation by introducing the normalized radius as $x = r$/R$_\mathrm{i}$. We assumed a power-law distribution for $C_f(x)$ because $C_f$ measures how the projected area of the outflowing gas changes relative to the projected area of the background stars \citep{martin09, steidel10}. If the outflowing clouds remain a constant constant size as they accelerate outward, then $\gamma = -2$ as the total area increases as $r^2$ \citepalias{chisholm16b}. However, most of the fitted profiles have $\gamma$ less than than $-2$ (see \autoref{tab:prof}), indicating that the outflows expand as they accelerate \citepalias{chisholm16b}.

To describe $\tau(v)$, we assumed that the absorption profile is broadened by the bulk motion of the outflow, which is called the Sobolev approximation \citep{sobolev}. The Sobolev approximation describes $\tau(v)$ as 
\begin{equation}
    \tau(v) = \frac{\uppi e^2}{\mathrm{mc}} f \lambda_\mathrm{0} n(v) \frac{dr}{dv} = \frac{\uppi e^2}{\mathrm{mc}} f \lambda_\mathrm{0} n(w) \frac{\mathrm{R}_\mathrm{i}}{v_\infty} \frac{dx}{dw} ,
    \label{eq:tau}
\end{equation}
where $\lambda_\mathrm{0}$ is the rest wavelength of the \siivp~1402\AA\ line, m is the mass of the electron, $w$ is the velocity normalized by the maximum velocity ($w = v/v_\infty$), and $dx/dw$ is the radial velocity gradient. The radial velocity gradient describes the acceleration of the outflow: a small $dx/dw$ indicates that the outflow velocity rapidly increases with radius. The radial velocity gradient is unknown, but we assumed that the velocity varies as a $\beta$-velocity law \citep{lamers} with
\begin{equation}
w(x) =  \left(1-\frac{1}{x}\right)^\beta .
\label{eq:beta}
\end{equation}
$\beta$-velocity laws are expected if a constant force accelerates the galactic outflow \citep{murray05}. \citetalias{chisholm16b} found that NGC~6090 has an acceleration profile with $\beta = 0.43$. This $\beta$-law is consistent with an outflow that is accelerated by an $r^{-2}$ force that is opposed by gravity. Most of the galaxies in the sample have $\beta$ values near 0.5 (\autoref{tab:prof}). The derivative of \autoref{eq:beta} defines the velocity gradient in \autoref{eq:tau}, and the inverse of \autoref{eq:beta} is used to transform radial relations, i.e. $C_f(x)$, into the observed velocity frame, i.e. $C_f(w)$.  

The final component of the $\tau(v)$ distribution is n$(v)$. This describes how the density changes with velocity and it is an important input for the photoionization models. We assumed that the \siiv radial density profile (n$_\mathrm{4}\left(x\right)$) is also a power-law such that 
\begin{equation}
\text{n}_\mathrm{4}\left(x\right) = \text{n}_\mathrm{4, 0} x^{\alpha} ,
\label{eq:denlaw}
\end{equation}
where n$_\mathrm{4, 0}$ is the initial \siiv density. If the galactic outflow is a spherical, mass-conserving flow, then $\alpha = -2$, but $\alpha$ is more negative than $-2$ if the geometry differs from spherical \citep{fielding} or if gas is removed from the outflow \citep{leroy15, chisholm16b}. 

The full system of equations for the observed $\tau(w)$ and $C_f(w)$ is
\begin{equation}
\begin{aligned}
\tau\left(w\right) &= \frac{\upi e^2}{\mathrm{mc}} f \lambda_\mathrm{0} \frac{\mathrm{R}_\mathrm{i}}{\mathrm{v}_\infty} \mathrm{n}_\mathrm{4, 0} x^\alpha \frac{dx}{dw}= \tau_0 \frac{w^{1/\beta-1}}{\beta(1-w^{1/\beta})^{2+\alpha}}\\
C_f(w) &=  \frac{C_f (\mathrm{R}_\mathrm{i})}{\left(1-w^{1/\beta}\right)^\gamma} ,
\label{eq:cfbeta}
\end{aligned}
\end{equation}
and we used {\small MPFIT} \citep{mpfit} to solve for the five free parameters ($\tau_0$, $C_f(\mathrm{R}_\mathrm{i})$, $\alpha$, $\gamma$, $\beta$; see \autoref{tab:prof}). The constant $\tau_0$ describes the maximum optical depth of the line, but it also allows us to calculate R$_\mathrm{i}$ from \autoref{eq:tau} as:
\begin{equation}
\mathrm{R}_\mathrm{i} = \frac{\mathrm{mc}}{\uppi e^2} \frac{1}{f \lambda_\mathrm{0}} \frac{\tau_0 \mathrm{v}_\infty}{ \mathrm{n}_\mathrm{H,0} \chi_\mathrm{Si IV} \mathrm{(Si/H)}} ,
\label{eq:tau0}
\end{equation}
where we replaced n$_{4, 0}$ with the initial total hydrogen density (n$_{\mathrm{H}, 0}$) times the \siiv ionization fraction ($\chi_\mathrm{Si IV}$) and the silicon-to-hydrogen abundance (Si/H). These parameters convert n$_{4,0}$ to n$_{\mathrm{H}, 0}$ and are derived in the photoionization modeling of \autoref{cloudy}.

Finally, in \citetalias{chisholm17}, we used these profiles to calculate the velocity-resolved mass outflow rate (\moutp$(w)$) as 
\begin{equation}
\begin{aligned}
    \dot{M}_\mathrm{o}(\mathrm{r}) &= \Omega C_{f}(\mathrm{r}) \mathrm{v}(\mathrm{r}) \rho(\mathrm{r}) \mathrm{r}^2 \\
    \dot{M}_\mathrm{o}(w)&= \Omega C_f(\mathrm{R}_\mathrm{i}) v_\infty \mu \mathrm{m}_\mathrm{p}  \mathrm{n}_\mathrm{H,0} \mathrm{R}_\mathrm{i}^2 \frac{w}{(1-w^{1/\beta})^{2+\gamma+\alpha}} ,
    \label{eq:mout}
\end{aligned}
\end{equation}
where $\Omega$ is the angular extent of the outflow  (we assume 4$\uppi$, or a spherical outflow) and $\mu \mathrm{m}_p$ is the mass per average nucleon (1.4 times the proton mass for standard  compositions). These values were discussed in detail in \citetalias{chisholm17} and the maximum \moutp$(w)$ values are tabulated in \autoref{tab:prof}. The \mout errors were calculated by creating 1000 iterations of \mout by varying the individual parameters in \autoref{eq:mout} by a Gaussian distribution centered on zero with a standard deviation equal to the parameter error derived by {\sc MPFIT}. These \mout values are used in \autoref{results} to explore the rate at which outflows remove metals from galaxies. 

\subsection{Ionization Modeling}
\label{cloudy}

\begin{figure*}
\includegraphics[width = \textwidth]{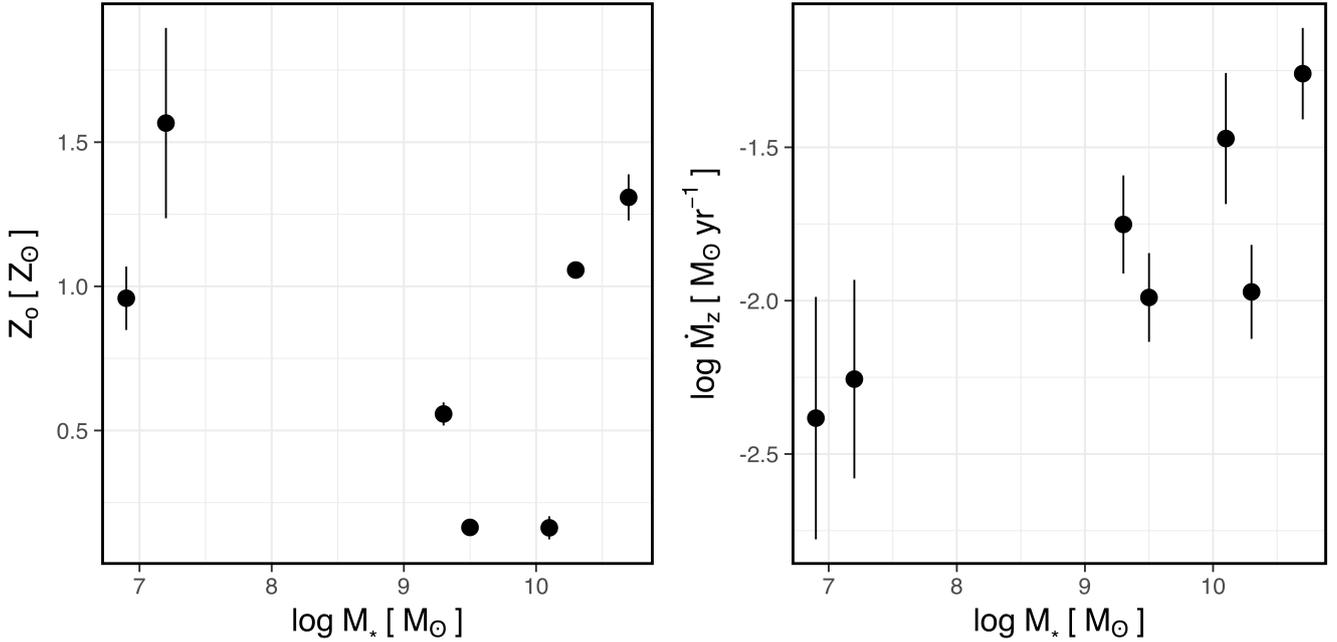}
\caption{The relation between the outflow metallicity (\zop; left panel) and the metal outflow rate ($\dot{\text{M}}_\text{Z} = \dot{\text{M}}_\text{o} Z_\text{o}$; right panel) with stellar mass (\mstarp). \zo does not vary with \mstarp, but \mz\ correlates with \mstar\ at the 2$\sigma$ significance level (\autoref{eq:metaloutflow}).}
\label{fig:z}
\end{figure*}

In the past two subsections we modeled the ionization source (the O and B stars within the COS aperture in \autoref{cont}) and the radial density profile (\autoref{proffit}). Here, we describe how we used the observed column densities to model the ionization structures and outflow metallicities (\zop) of the galactic outflows. 

To estimate the ionization structure, we created a large grid of {\small CLOUDY} photoionization models \citep[v13.03;][]{ferland} by varying the ionization parameter (U$ = n_\mathrm{ph}/n_{\mathrm{H}, 0}$) as $-2.75 < \mathrm{log(U)} < -1.5$, the density at the base of the outflow as $-1 < \mathrm{log(}n_{\mathrm{H}, 0}[\mathrm{cm^{-3}])} < 4$, and \zo as $0.01 < \mathrm{Z}_\mathrm{o} < 2.5$~Z$_\odot$. We created each of these models using the best fit {\small STARBURST99} stellar continuum model as the ionizing source, a spherical geometry with the fitted radial density profile, and the observed $C_f(\mathrm{R}_\mathrm{i})$ as the covering fraction. Importantly, the steeply declining \no\ distributions (see $\alpha$ in \autoref{tab:prof}) mean that the geometries are nearly thin shells, but this was not assumed.  We included {\small CLOUDY}'s default \ion{H}{ii} abundances, which have an Orion nebular dust distribution, to account for depletion of gas on to grains \citep{baldwin1991}. The abundances were multiplied by constant factors of \zo to enrich, or deplete, the outflow. This \ion{H}{ii} abundance set defines the individual abundances and we did not rescale the individual abundances (like the Si-to-S or C-to-O ratios). These abundances have 12+log(O/H)$_\odot$ = 8.60, and we rescaled all of the {\small CLOUDY} derived metallicities onto the abundance scale of \citet{asplund}. We also included a cosmic ray background \citep{indriolo} because our stopping condition is 500~K. For each combination of U, $n_{\mathrm{H}, 0}$, and \zop, we tabulated the integrated \oip, \siiip, \siip, and \siiv column densities predicted by {\small CLOUDY}.

Once the model column densities were tabulated, we determined the best fit parameters using a Bayesian approach. We considered each model equally likely (a flat prior) with a likelihood function given by
\begin{equation}
L \propto e^{-\chi^2/2} ,
\end{equation}
where $\chi^2$ is the chi-squared function derived using the integrated \oip~1302\AA, \siiip~1304\AA, \siip~1250\AA, and \siivp~1402\AA\ column densities, which were corrected for non-unity covering fraction (see \autoref{tab:column}). The five massive galaxies have similar covering fractions for all of their transitions, but the two low-mass galaxies have different \siii and \siiv covering fractions (see \autoref{fig:s42}). For these two galaxies, we measured the \siii covering fraction using the \siiip~1260\AA\ line. We then corrected the \oip, \siiip, and \sii column densities  with these \siii covering fractions (1~Zw~18 and SBS~1415+437 have \siii covering fractions of 0.45 and 0.60). 

The four transitions used in the photoionization modeling are fairly weak: the median of the \siiv doublet ratio is 1.7 for the sample, while the \siip~1250\AA\ and \oip~1302\AA\ line have oscillator strengths of of 0.006 and 0.05, respectively. Importantly, \oi has a similar ionization potential to \hi, such that the \oi ionization state is locked to \hi\ through charge exchange. Therefore, the \oi column density is a strong proxy of neutral gas.  We then marginalized the likelihood function over nuisance parameters and calculated the expectation values and standard deviations of U, \no, and \zo from these probability density functions. These values are listed in \autoref{tab:ion}.

We then produced a {\small CLOUDY} model for each galaxy with these expectation values. These best fit {\small CLOUDY} models describe the ionization structure of the outflows, given the observed column densities. From these {\small CLOUDY} models we recovered the ionization fractions ($\chi$), silicon-to-hydrogen gas-phase abundances (Si/H), and the \ion{H}{i} and the total H column densities (see \autoref{tab:ion}). In \autoref{cloudy_obs}, we explore how these modeled ionization properties relate to observable properties not included in the modeling, as well as how the observables change over the full {\small CLOUDY} grid.

\section{RESULTS}
\label{results}
This paper focuses on new measurements of the outflow metallicity (\zop) to test whether galactic outflows shape the mass-metallicity relationship. Therefore, the scaling of \zo with \mstar is our fundamental result. \zo does not vary with \mstar (left panel of \autoref{fig:z}), instead \zo scatters about $1.0\pm0.6$~Z$_\odot$.  

Meanwhile, the metal outflow rate (\mzp~=~\zop$\times$\moutp) increases with \mstar (right panel of \autoref{fig:z}). We find  a 2$\sigma$ significant correlation (p-value < 0.02) with
\begin{equation}
\dot{\mathrm{M}}_\mathrm{z} = \left(0.021 \pm 0.005\right) \mathrm{M}_\odot~\mathrm{yr}^{-1} \left(\frac{\mathrm{M}_\ast}{10^{10} \mathrm{M}_\odot}\right)^{0.23 \pm 0.06} .
\label{eq:metaloutflow}
\end{equation}
This relationship needs statistically significant numbers to confirm (\autoref{further}), but it provides a scaling for the metal outflow rate of the galaxies within our sample. We discuss the implications for these trends and how they shape the MZR, in \autoref{mzrel}. 
\section{DISCUSSION}

\subsection{Comparison to previous results and simulations}
\label{prev}
\subsubsection{Previous observations}
The local galaxies in our sample are well-studied, and previous studies have estimated similar properties as our photoionization models. For instance, log($N_\mathrm{HI}[\mathrm{cm}^{-2}]$) was estimated for MRK~1486 \citep{duval} and Haro~11 \citep{rivera-thorsen17} by modeling the \siii absorption lines and the \lya\ emission profiles. These previous studies found log($N_\mathrm{HI}[\mathrm{cm}^{-2}]$) of 19.5 and $> 19.0 \pm 0.2$, respectively. These independent values agree with the estimated log(\nhi) found from our {\small CLOUDY} modeling (see \autoref{tab:ion}).  

Other studies carefully measured the metallicity from the low metallicity galaxy 1~Zw~18 using the observed \lya\ profile and the sulphur absorption lines to determine a metallicity of $\sim0.02$~Z$_\odot$ in the ionized gas and $\sim0.03$~Z$_\odot$ in the neutral gas \citep{lebouteiller}. This metallicity is substantially lower than the $1.6\pm0.3$~Z$_\odot$ that we derived using the {\small CLOUDY} models, and similar to \zism. Similarly, \citet{james} estimated log(\nhi)$ = 21.1$ from SBS~1415+437. The \ion{H}{i} column densities and metallicities measured here are significantly different from these studies.

The \lya\ regions for 1~Zw~18 and SBS~1415+437 are extremely difficult to interpret due to the presence of damped Milky Way absorption \citep{kunth}, geocoronal emission, absorption from high-velocity clouds \citep{hvc1, hvc2}, \lya\ plus nebular emission from the background galaxy \citep{lebouteiller}, and stellar continuum features. Importantly, the broad \ion{N}{v}~1240\AA\ P-Cygni feature and stellar \ion{H}{i} absorption can be interpreted as damped Ly$\alpha$ wings. \citet{james} and \citet{lebouteiller} carefully account for many of these features, but \citet{james} do not fit the stellar continuum and \citet{lebouteiller} fit the stellar continuum after measuring the \ion{H}{i} column density. Not accounting for the stellar continuum may over-estimate the \ion{H}{i} column density and under-estimate the metallicity.

Additionally, we accounted for partial covering when determining the metal column densities while \citet{lebouteiller} and \citet{james} assumed a unity covering fraction. We inferred the covering fractions from the observed absorption profiles, thus the covering fractions should be accounted for when calculating column densities. Since both 1~Zw~18 and SBS~1415+437 have small \siiv covering fractions (see \autoref{tab:prof}), assuming the lines are fully covered systematically decreases the estimated column densities from Voigt fits. This may have led to the large differences between the measured metallicities.

Further, the photoionization models of \citet{lebouteiller} are only observationally constrained by the low-ions (up to \siiip). The authors find a factor of five difference in the metallicities derived in the neutral and ionized Si zones. By not observationally constraining the ionization models with high-ions, it is possible that the ionization corrections of the high-ions could be even larger. Our ionization models suggest that 58\% of the Si from 1~Zw~18 is in the \siiii or \siiv transitions; not including higher ionization phases may under-predict the total amount of outflowing metals.

{\small CLOUDY} models with 0.01~Z$_\odot$ under-predict the \siiv column densities by over an order of magnitude for all U and \no\ combinations, even though they allow for the large log(\nhi) values found by previous studies (the {\small CLOUDY} models exceed log(\nhi) = 26). In \citetalias{chisholm16}, we also found that very low \zo were incompatible with the observed ionization structure of the outflows (fig.~14 of that paper). Further, the low metallicity {\small CLOUDY} models produce larger \siivp/\siii column density ratios than are observed (see \autoref{cloudy_obs} and \autoref{fig:cloudy_mod} below). 

Another possibility is that these two low-mass galaxies have substantial amounts of low-metallicity gas at zero-velocity, and the strong damped \lya\ absorption traces this systemic gas which has the ISM metallicity. The metal absorption lines are from an enriched galactic outflow, while the \lya\ from the outflow is blended with the systemic damped \lya. The deepest portion of the low-ionization metal absorption lines from 1~Zw~18 are blueshifted, relative to the 21~cm line, by 15~\kms \citep{lebouteiller}, while the equivalent width weighted velocity of the \siivp~1402\AA\ line is blueshifted by $-169\pm15$~\kms (see \autoref{fig:s42}). The low-velocity \hi\ absorption is easily blended with zero-velocity gas, making it challenging to disentangle the relatively weak outflow--as traced by the metal lines--from the strong systemic \lya\ absorption.  Zero-velocity contamination is a significant problem for outflow studies \citep{weiner, chen10, rubin13}, and it could play a large role in the discrepancies between the metallicities measured here and previous studies. 

To this end, \citet{lelli} observed 21~cm emission from 1~Zw~18 with the VLA. These authors modeled the \ion{H}{i} emission maps as a disk plus a radial flow with a best fit velocity of 15~\kms \citep[the same as the low-ionization absorption lines observed by][]{lebouteiller}. Interestingly, the \hi\ disk is not centered on the star-forming region probed by the COS aperture, possibly suggesting that the \hi\ and star formation are not coincident. Once they removed the disk component, \citet{lelli} found a residual log(\nhi[cm$^{-2}$])~$=19.6$, corresponding to outflowing gas (their fig. 8), consistent with the ionization modeling here. Therefore, it is plausible that the metal absorption lines from the two low-mass galaxies trace a metal-enriched outflow, while the \lya\ absorption traces a large-scale, low-metallicity, zero-velocity disk.

Other studies have measured outflow metallicities, but not in the warm photoionized outflow phase that we observed. \citet{martin02} fit the integrated X-ray spectrum of a low-metallicity (\zism~$=0.2$~Z$_\odot$; 12+log(O/H) = 8.0; log(\mstarp)~$=7.8$) galaxy, NGC~1569, with a four component model: a disk, a power-law, an absorbed thermal component, and an unabsorbed thermal component. Both thermal components were treated as the outflow. The authors minimized the fit to the X-ray spectrum when the $\alpha$ elements (like Si and O used here) had \zop~$=1$~Z$_\odot$,  {but \zo is not tightly constrained due to a broad \zo distribution (see their fig. 19). Regardless, the solar metallicity measurement of the \textit{hot} outflow is}  in agreement with the median outflow metallicity we measured  {for the warm outflow phase} in \autoref{results}, and 5 times larger than \zism. 

X-ray studies of 10$^7$~K plasma from the local starbursting galaxy M~82 find super-solar metallicities that are >$3$~Z$_\odot$ \citep{strickland09}. This large \zo likely arises because the large-scale X-ray emission traces pure supernovae ejecta, whereas the COS metal absorption lines trace a mixture of ISM gas and supernovae ejecta. Removing ISM gas drives the MZR, thus tracing the metallicity of the gas being removed is crucial for determining which physical process creates the MZR. We return to the idea of mixing the two phases in \autoref{highz}.

\subsubsection{Simulated metal outflow rates}
\label{sim}

The \mz measures how rapidly outflows remove metals from galaxies. Simulations usually tune \mz to match the observed MZR (see the discussion in \autoref{mzrel}). However, recent simulations from the Feedback In Realistic Environments simulations \citep[FIRE;][]{hopkins14} enrich gas according to the supernovae yields \citep{woosley95}, while driving feedback by injecting energy and momentum into the gas according to stellar population synthesis models. The metal production and feedback processes successfully reproduce the MZR at $z = 0-3$ \citep{ma}. \citet{muratov17} measure \mz at 0.25~R$_\mathrm{vir}$ for a small sample of these simulated galaxies. Two galaxies in particular compare well with our sample because they have \mstar of $2\times10^9$ and $6\times 10^{10}$~M$_\odot$ at $z = 0$ (their m11 and m12i simulations;  {the values come from the peaks in the left panel of their fig.~1}). These galaxies have \mz peak values between $0.01-0.04$~\sfr for the low-mass galaxy, and $0.05-0.11$~\sfr for the high-mass galaxy (after converting to the solar metallicity conventions used here). These two simulated galaxies are comparable to MRK~1486 (comparable to m11, the low-mass FIRE galaxy) and NGC~6090 (comparable to m12i, the high-mass FIRE galaxy), which have \mz of $0.02\pm0.01$ and $0.06\pm0.02$~\sfrp, respectively.  {The low-mass FIRE galaxy, m10, has \mstar at $z = 0$ three times less massive than the lowest mass galaxy in our sample (2.3$\times10^{6}$~M$_\odot$), which makes it challenging to directly compare to our sample. {However, if we extrapolate} the fitted relationship between \mz and \mstar in \autoref{eq:metaloutflow}, we predict a \mz of 0.0031$^{+0.0032}_{-0.0016}$~\sfr for a galaxy with \mstar of 2.3$\times10^{6}$~M$_\odot$, consistent, within the large errors, with the maximum of the \mz range of 0.0002-0.0016~\sfr for m10.}  The simulated \mz values are surprisingly consistent with our observations. In \autoref{mzrel} we discuss the implications of the \mz values and why they determine whether simulations reproduce the observed MZR.

\subsection{Shaping the mass-metallicity relation with galactic outflows}
\label{mzrel}

\begin{figure*}
\includegraphics[width = \textwidth]{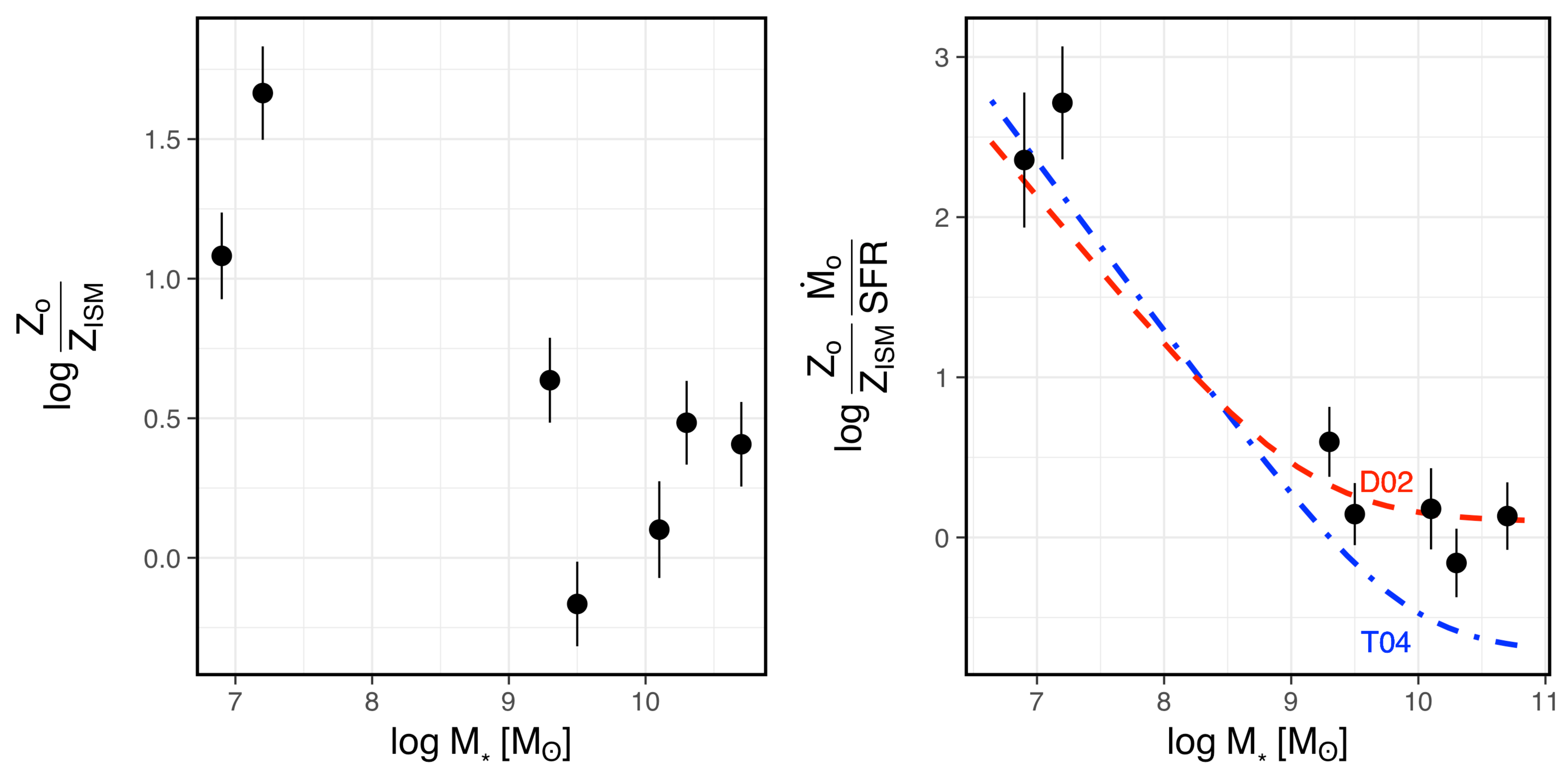}
\caption{Relationships that shape the mass-metallicity relation. {\it Left  panel:} Outflow metallicity (\zop) normalized by the gas-phase metallicity of the galaxy (\zism) versus the stellar mass of the galaxy (\mstarp). {\it Right panel:} The logarithm of the metal-loading factor, the ratio of metals removed by the galactic outflow to metals locked into stars, versus the logarithm of \mstarp. The metal-loading factor inversely scales with \mstar at the 3$\sigma$ significance level (p-value < 0.001). Included in the right panel are two analytic predictions of the metal-loading factor required to match the observed MZR from \citet{denicolo} (red dashed line) and \citet{tremonti04} (blue dot-dashed line), as fit by \citet{peeples11}. The observed metal-loading factors closely follow the analytic relation required to reproduce the observed \citet{denicolo} MZR. }
\label{fig:zout}
\end{figure*}

Here we explore the implications that the observed metal outflows have on the stellar mass gas-phase metallicity relation (MZR). The left panel of \autoref{fig:z} indicates that \zo does not vary with \mstarp, rather \zo scatters about a median of 1.0~Z$_\odot$. However, when \zo is normalized by the observed ISM gas-phase metallicity (Z$_\mathrm{ISM}$; left panel of \autoref{fig:zout}), the outflow metallicities from six of the seven galaxies are more metal-enriched than their ISM metallicities. Even at  {the upper-end of the mass range of our sample} (log(\mstarp[M$_\odot$)~$>9$), the median \zo is 2.6 times larger than \zism. The outflows of the lowest mass galaxies carry out even more metals as their outflow metallicities are 12 and 46 times larger than their ISM metallicities. The observed outflows are metal-enriched.

\citet{dalcanton} analytically studied three possible origins of the MZR: inflows of low-metallicity gas, low-metallicity outflows, and metal-enriched outflows. The author finds that outflows with \zop/\zism~>~1 are the only way to decrease the metallicity of low-mass galaxies to the observed levels  \citep[a similar conclusion was found by][]{lu}. However, \citet{dalcanton} did not consider the possibility that outflows remove more mass in lower mass galaxies than in higher mass galaxies. \citet{peeples11} introduced the metal-loading factor, which describes how efficiently galactic outflows remove metals relative to how efficiently star formation retains metals. Numerically, this is given as
\begin{equation}
\zeta = \frac{Z_\mathrm{o}}{Z_\mathrm{ISM}} \frac{\dot{\mathrm{M}}_\mathrm{o}}{\mathrm{SFR_\text{COS}}} ,
\label{eq:zeta}
\end{equation}
where we used \sfrc\ in \autoref{eq:zeta} because we assumed that the local star formation energy and momentum accelerates the outflow (see \autoref{sample}). The numerator in \autoref{eq:zeta} is the rate at which outflows remove metals and the denominator is the rate at which star formation retains metals in new stars. In essence, what drives the MZR is the competition between the ejection and retention of metals. This tension is the metal-loading factor (right panel of \autoref{fig:zout}). The metal-loading factor strongly anti-correlates with \mstar (3$\sigma$, p-value < 0.001, relation; Pearson's $r$ of -0.97).

To better understand this competition, consider how the metallicity of a galaxy changes with time \citep{larson74, tinsley80, dalcanton, finlator08, peeples11, lilly13, zahid, lu}.  Galaxies accrete low-metallicity gas from the IGM (with a metallicity of Z$_\mathrm{IGM}$) at a rate $\dot{M}_\mathrm{acc}$. This low-metallicity gas collects in the gravitational potential of the galaxy and forms stars. These stars produce metals (at a yield $y$) and instantly return a fraction, $R$, of these metals to the ISM through stellar winds or supernovae. The same stellar winds and supernovae inject energy and momentum into the newly enriched gas to drive a galactic outflow which removes a mass of metals at a rate of $\dot{M}_z= Z_\mathrm{o}\times\dot{M}_\mathrm{o}$. This approach, often called the leaky box (or bathtub) model, describes the change in \zism\ as a differential equation:
\begin{equation}
\frac{\mathrm{dZ}_\mathrm{ISM}}{\mathrm{dt}} = -\mathrm{Z}_\mathrm{ISM} \text{SFR} +\mathrm{Z}_\mathrm{IGM} \dot{M}_\mathrm{acc} -Z_o \dot{M}_o + y (1-R) \text{SFR} .
\label{eq:mmt}
\end{equation}
Assuming that $Z_\mathrm{IGM} = 0$, $y$ is constant, and that metals are instantly recycled back to the ISM, \citet{peeples11} solved this differential equation as
\begin{equation}
\mathrm{Z}_\mathrm{ISM} = \frac{y}{\zeta +\alpha \frac{M_g}{M_\ast} +1} ,
\label{eq:peep}
\end{equation}
where $\alpha$ is a constant of order 1 and ${M_g}$/${M_\ast}$ is the gas-to-stellar mass fraction of the galaxy. \autoref{eq:peep} resembles the closed box model of the MZR (\autoref{eq:closed}), but includes the loss of metals by galactic outflows. 

Assuming that $y$ is constant, \autoref{eq:peep} shows that the MZR is completely determined by the scaling of the metal-loading factor and the gas mass fraction. If galaxies accrete metal rich gas ($Z_\mathrm{IGM} \ne 0$), possibly due to a galactic fountain or an enriched IGM, then the denominator of \autoref{eq:peep} must include the accretion loading factor ($\frac{\mathrm{Z}_\mathrm{IGM}}{\mathrm{Z}_\mathrm{ISM}} \frac{\dot{M}_\mathrm{acc}}{\mathrm{SFR}}$), but we ignore this effect because the IGM is assumed to have a negligible metallicity.

The scaling of the metal-loading factor with the circular velocity (\vcircp; a proxy of the stellar mass) is fit with a power-law as
\begin{equation}
\zeta = \left(\frac{\mathrm{v}_0}{\mathrm{v}_\mathrm{circ}}\right)^{b}+\zeta_0 .
\label{eq:peep_zeta}
\end{equation}
The shape of the metal-loading factor relation has two portions: (1) a low-mass portion below the turn-over mass (v$_0$) where the metal-loading factor rapidly increases with decreasing \mstar as a power-law with exponent $b$, and (2) a high-mass portion, above v$_0$, that has a constant metal-loading factor ($\zeta_0$). We fit \autoref{eq:peep_zeta} to the observed metal-loading factors and find v$_0=91\pm22$~\kmsp, $b=3.4\pm0.7$, and $\zeta_0=0.74\pm0.39$.

\citet{peeples11} placed the scaling of eight different MZRs from the literature into \autoref{eq:peep}, along with the observed $M_g/M_\ast$ scaling from three combined samples \citep{mcgaugh05, leroy08, west09, west10}, and fit for the power-law metal-loading factor required to reproduce the observed MZRs. In the right panel of \autoref{fig:zout}, we over-plot their results for the \citet{tremonti04} (blue dot-dashed line) and \citet{denicolo} (red dashed line) relations. \citet{denicolo} calibrated the [\ion{N}{ii}]/H$\alpha$ ratio as a metallicity indicator from a sample of galaxies with a mixture of T$_e$ based oxygen abundances and photoionization model based abundances.   Meanwhile, \citet{tremonti04} used a Bayesian approach to estimate metallicities using all of the available strong optical emission lines and a grid of photoionization models. The differences between the two calibration systems--especially at the high-mass end--reflect the large uncertainties of different metallicity calibrations \citep{kewley08}. Other MZRs have similar shapes as the two presented here \citep[see fig. 6 of][]{peeples11}, and the fits to the observed metal-loading factors are broadly consistent with the predictions from \citet{peeples11}. The observed metal-loading factors agree better with the metal-loading factors required to reproduce the \citet{denicolo} relation. This is not surprising because we used T$_e$ based \zism\ values, similar to the \citet{denicolo} calibration.

The low-mass (log(\mstarp)~$=7.8$) galaxy from \citet{martin02} also resides on the curves of \autoref{fig:zout}. With a \zop/\zism~$=5$ and an estimated mass-loading factor of 9  {for the hot outflowing component}. A metal-loading factor of 45 (log$\zeta = 1.65$) resides between both the \citet{denicolo} and \citet{tremonti04} curves in \autoref{fig:zout}. This suggests that the metal-loading factor increases monotonically within the \mstar gap in \autoref{fig:zout}, but future observations are required to confirm this (see \autoref{further}). 

In \autoref{eq:peep}, the declining $M_g/M_\ast$ and the observed metal-loading factors completely describe the shape of the observed MZR. At low-masses (log($M_\ast$[M$_\odot$])$~< 10$),  $M_g/M_\ast$ and the metal-loading factors are greater than 1, and both decrease with increasing \mstarp. Therefore, \zism\ increases with increasing \mstarp. However, for log(\mstarp[M$_\odot$])~>~10, $M_g/M_\ast$ is much smaller than 1 and the metal-loading factor flattens to a constant value. This reduces the denominator in \autoref{eq:peep} to $\zeta_0 + 1$, and \zism\ becomes a constant equal to
\begin{equation}
\text{Z}_\text{flat} =\frac{y}{\zeta_0+1} . 
\label{eq:zflat}
\end{equation} 
Consequently, the observed Z$_\text{flat}$ from various MZRs provides $y$ estimates. Using Z$_\text{flat}$ of 0.006 and 0.014 from the \citet{denicolo} and \citet{tremonti04} MZRs, we estimate $y$ to be 0.010 and 0.024. These values agree with the range of oxygen yields $y~=~0.007-0.038$ recently calculated by \citet{vincenzo}. This range depends on the IMF, \zs, and the high-mass cut-off. The implied $y$ range of 0.007--0.038 is most similar to the Salpeter and Kroupa oxygen yields \citep{salpeter, kroupa}. Conversely, the Chabrier IMF \citep{chabrier} predicts substantially larger $y$ than implied by the flat portions of the MZR.

Since \zop/\zism\ does not scale with \mstarp, the mass-loading factor (\moutp/SFR) alone determines the metal-loading factor. In \citetalias{chisholm17}, we showed that the mass-loading factor decreases with \mstar because the gravitational potential increases more rapidly with \mstar than the momentum supplied by star formation \citepalias[fig.~5 of][]{chisholm17}. This declining efficiency of star formation driven outflows reduces the metal-loading factor and flattens the MZR. The observed physics of driving galactic outflows shapes the stellar mass-metallicity relation.

The scatter of the MZR is small \citep[0.1~dex; ][]{tremonti04}. This may imply that the metal-loading factors and gas mass fractions do not strongly vary at a given \mstar (see \autoref{eq:peep}). However, recent \ion{H}{i} observations find a strong correlation between the \ion{H}{i} gas mass and 12+log(O/H) at fixed \mstarp, implying that the scatter in the MZR is tied to the \ion{H}{i} gas mass \citep{brown}. This suggests that the metal-loading factors either have a small scatter, or that the \ion{H}{i} gas mass and metal-loading factors are related. Future observations of both the metal-loading factors and the gas masses are required to break this degeneracy to determine what causes the scatter in the MZR.

\subsubsection{Must outflows from low-mass galaxies be metal-enriched to match the MZR?}
\label{highz}

The lowest mass galaxies in our sample have \zo exceeding their \zism\ by factors of 10-50 (left panel of \autoref{fig:zout}). This starkly contrasts the more massive galaxies in the sample which have Z$_\text{o}~\sim~{\rm 2Z}_\text{ISM}$. Are these highly enriched outflows required to reproduce the observed MZR?

If \zo is the only observed property that changes, then the derived metal-loading factors do not change. This counter-intuitive fact arises because the metal-loading factor is a product of \zo times \moutp. Changes in \zo also propagate to \mout by a factor of \zop$^{-1}$ from R$_\mathrm{i}^{2}$ in \autoref{eq:mout} (where a factor of Z$_\mathrm{o}^{-2}$ comes directly from \zo in \autoref{eq:tau0}, and a factor of Z$^{+1}_\mathrm{o}$ comes from $\chi_\mathrm{SiIV}$ in \autoref{eq:zfrac}).  Consequently, our measured \mout changes as $Z_\mathrm{o}^{-1}$, but the metal-loading factor remains {\it constant} because the metal-loading factor is the product \moutp$\times$\zop. 

For example, if the \zo of SBS~1415+437 hypothetically decreased from the modeled 0.96~Z$_\odot$ to the \zism\ of 0.08~Z$_\odot$, then \mout would increase from 0.3~\sfr to 3.6~\sfrp. This would increase the mass-loading factor from 19 to 225, but the {\it metal}-loading factor would remain 225. Therefore, if SBS~1415+437 did not drive an enriched outflow, then the outflow must remove {\it 12 times more} mass to match the observed MZR. Consequently, metal enriched outflows are {\it not} required to match the MZR, but if the outflows are not metal-enriched then the mass outflow rate must be sufficiently large enough to produce the required metal-loading factor. 

The outflow metallicity is an important new constraint for galaxy simulations, and may help to produce more realistic simulations. Typically, cosmological simulations drive galactic outflows with \zop~$ = $~\zism\ \citep{finlator08, dave11, schaye15, christensen, muratov17} or with $\mathrm{Z}_\mathrm{o} = 0.4 \mathrm{Z}_\mathrm{ISM}$ \citep{vogelsberger13}. The mass-loading factor corresponding to \zop~=~\zism\ for SBS1415+437 is 225, which is more consistent with the mass-loading factor of 140 predicted by the FIRE simulations \citep[eq. 4 in][]{muratov} than the mass-loading factor of 19 we estimated in \citetalias{chisholm17}. Thus, a simulation with \zop~=~\zism\  would drive 12~times more mass out of the galaxy than suggested by our observations.   Simulations often have troubles producing starburst galaxies \citep{sparre15}, possibly because the simulations remove too much mass in order to match the observed MZRs. If galactic outflows are metal-enriched, then cosmological simulations that reproduce the observed MZR may drive too massive of galactic outflows and deplete their gaseous reservoir too rapidly.

\subsubsection{How are outflows enriched?}
\label{drive}

\begin{figure}
\includegraphics[width = 0.5\textwidth]{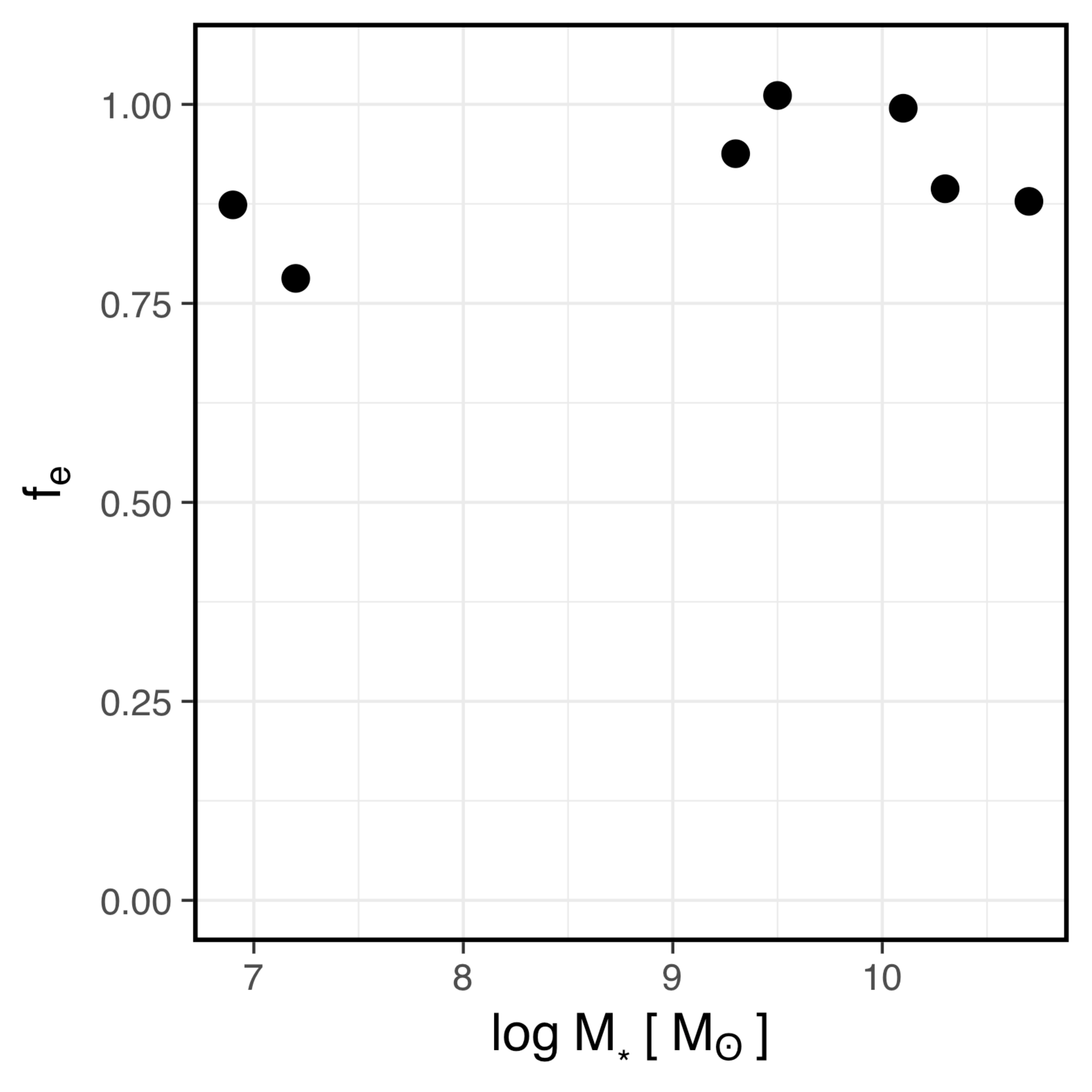}
\caption{The entrainment fraction ($f_e$; \autoref{eq:fe}) versus stellar mass (\mstarp). The entrainment fraction is the fraction of the outflow that is entrained ISM versus pure supernovae ejecta. An $f_e$ of 1.0 indicates that the outflow is entirely swept up ISM, while a lower $f_e$ indicates that the outflow is enriched by supernova ejecta. Outflows are largely swept up ISM, while outflows from low-mass galaxies have moderate enrichment.}
\label{fig:fentrain}
\end{figure}
 
The outflow metallicity provides clues about the origins of the outflowing gas. A \zop/\zism~$=1$ would indicate that the outflow is entirely entrained ISM. Meanwhile, elevated \zop/\zism\ ratios imply that outflows are enriched by recent star formation (supernovae or stellar winds). Supernova ejecta have Z$_\mathrm{ej} \sim 0.1 = 7~Z_\odot$ \citep[][]{woosley95}, and this large supply of metals can mix with the ambient ISM to increase \zop. We observe that \zo is larger than \zism\ for six of the seven outflows (left panel of \autoref{fig:zout}),  {and statistically consistent with being equal to \zism\ for one galaxy}, implying that the observed outflows are entrained ISM with some metal enrichment.

We estimate the fraction of the outflow that is pure ISM.  The "entrainment fraction" \citep[$f_\mathrm{e}$;][]{dalcanton, peeples11} is the fraction of the outflow that is purely swept up ISM, and is given as
\begin{equation}
f_e = \frac{Z_\mathrm{o}- Z_\mathrm{ej}}{Z_\mathrm{ISM} - Z_\mathrm{ej}} ,
\label{eq:fe}
\end{equation}
where we used $Z_\mathrm{ej} = 0.1$. An $f_\mathrm{e}$ of 1.0 indicates that the outflow is entirely swept up ISM, while an $f_e$ of zero indicates that the outflow is entirely supernova ejecta. Values between these extremes indicate that the ISM is a mixture of supernova ejecta and ambient ISM. High-mass galaxies have $f_\mathrm{e}$ values between 88 and 100 per cent (\autoref{fig:fentrain}); these outflows are nearly pure ISM. Even at the lowest stellar masses, 78 per cent of the outflowing gas from 1~Zw~18 is entrained ISM and only 22 per cent is pure supernova ejecta. Galactic outflows are largely entrained ISM with a small fraction of supernova enrichment. 

One possible origin of the $f_e$ behavior is that \zo of high-mass galaxies saturates as \zo approaches $y$.  All outflows may have a flat \zop$\sim1$~Z$_\odot$, regardless of \zism\ or \mstar (see \autoref{fig:z}). A flat \zo causes $f_e$ to increase with increasing \zism\ until it reaches 1.0 at \zism~$=1$~Z$_\odot$ where it remains constant. A fixed \zo could arise if $y$ is fixed with \zism. This is compatible with current model yields which show that $y$ only changes by a factor of 1.06 from \zs\ of 0.05 to 2.0~Z$_\odot$, but varies strongly with different IMFs \citep[by a factor of $4-5$, see discussion above;][]{vincenzo}. Thus, an evolving IMF or an increasing $\zeta_0$ (removing more metals) are the most likely determinants of Z$_{\rm flat}$ (see \autoref{eq:zflat}), which leads to the lower Z$_{\rm flat}$ that are observed for higher redshift galaxies \citep{savaglio05, erb06}.

Another possibility is that the observed outflows are enriched and driven differently in high and low-mass galaxies. These different physical driving mechanisms produce different \zop/\zism\ ratios. The smaller \zop/\zism\ of high-mass galaxies implies that these outflows largely remove ambient ISM material, which is expected if a supernova blast wave does not break out of the ISM, and instead "snowplows" ISM outward.  Meanwhile, low-mass galaxies do not fully contain the supernovae blast wave, rather the ejecta rips holes in, and mixes with, the ambient ISM. 

Hints of this dichotomy are seen in the \siii and \siiv line profiles (\autoref{fig:s42}). \siii and \siiv are co-moving in high-mass galaxies, as expected in an outwardly expanding snowplow phase. Meanwhile, the \siii in lowest mass galaxies extends to redder velocities than the \siivp. This indicates that high-ionization gas in the outflows of low-mass galaxies extends to higher velocities than the low-ionization gas. Unlike the high-mass galaxies, the \siii of the two low-mass galaxies also has a slightly larger $C_f$ than the \siivp. If $C_f$ decreases with $r$ (\autoref{eq:cf}), this suggests that the \siii originates at smaller radii than the \siiv. This scenario is expected if low-ionization gas is heated as it accelerates, as expected if a supernova ripped a hole in the ambient ISM \citep{chisholm18}. 

\subsection{ {An examination of our assumptions}}
 {Here we discuss the implicit assumptions associated with our stellar continuum fitting, physical outflow model, photoionization model, chemical evolution model, and host galaxy property calculations have on determining the role galactic outflows play in shaping the MZR.}

 {In \autoref{cont} we modeled the observerd stellar continuum as a linear-combination of instantaneous bursts of fully-theoretical single-aged \textsc{starburst99} models. This modeling determines the number of ionizing photons and the hardness of the ionizing spectra, which we used in the \textsc{cloudy} photoionization modeling to determine the ionization structure and outflow metallicity. In Papers~\textsc{I}--\textsc{III} we demonstrated that the linear-combination of stellar continuum models reproduces the observed massive star features like the \ion{N}{V} and \siiv P-Cygni stellar wind features, however the assumed IMF and stellar atmospheric models impact the number of ionizing photons produced by a single burst \citep{claus99}. For instance, binary stellar populations have recently been suggested to have a harder ionizing spectra with more ionizing photons, which may reproduce observed features like \ion{He}{ii} emission \citep{eldridge, stanway}. The two galaxies in the sample with \ion{He}{ii}~1640\AA\ coverage (1~Zw~18 and NGC~7714) have mixed \ion{He}{ii}~1640\AA\ results. 1~Zw~18 has strong \ion{He}{ii}~1640\AA\ \citep{lebouteiller} and optical \ion{He}{ii}~4363\AA, although models show that the \ion{He}{ii}~4363\AA\ emission is even stronger than predicted by binary stars alone \citep{kerig}. Meanwhile, NGC~7714 does not have detected \ion{He}{ii}~1640\AA\ emission, illustrating that the  hardness of the ionizing continua varies from galaxy-to-galaxy. Whether this is solely due to stellar metallicity affects, or requires a binary stellar population is still an unsolved problem.}

 {In \autoref{proffit} we fit for how the density and covering fraction scales with radius as well as the radial extent of the outflows. These parameters describe the acceleration and mass outflow rate. While we assume a physically motivated general form for the acceleration, density, and covering fraction; the exact scaling is observationally driven because we fit the parameters from the observed line profiles. This allows for galaxy-to-galaxy geometrical variations based on the observed outflow line profiles.  For instance, the density profile sharply decreases with increasing radius, implying that galactic outflows are thin, but not arbitrarily thin, shells. The statistical uncertainties of each fitted parameter are included in the \mout uncertainties, which incorporates the fitted geometrical uncertainties into the \mz uncertainties. As discussed in \citetalias{chisholm17}, this is especially apparent in low-mass galaxies where narrow line profiles make it challenging to fit the line profile (see \autoref{tab:prof}).}

 {The photoionization modeling in \autoref{cloudy} assumes that the observed phases have similar physical conditions and are in photoionization equilibrium. Galaxies with stellar mass greater than 10$^{9}$~M$_\odot$ appear to satisfy these criteria {because the low and high-ionization lines have similar profiles} (\autoref{fig:s42}), but lower mass galaxies may not. As discussed in \autoref{drive}, this may indicate that there is unaccounted for physics in the production and acceleration of galactic outflows arising from low-mass galaxies, and, if confirmed by future observations, may lead to the generation of more realistic galactic outflows in low-mass galaxies.}

 {In \autoref{cloudy} we also assumed that the relative abundances do not vary from galaxy-to-galaxy. We used the \ion{H}{ii} abundance set in \textsc{CLOUDY} to set the relative abundance. Observations do not detect a trend in the relative abundances of $\alpha$-elements, like Si/O or S/O that we use here,  with metallicity \citep{garnett95, izotov99}. This suggests that the assumption of constant relative abundance does not drastically impact the ionization models.}

 {One way to assess the impact of our assumed outflow model is to compare our \mout values to values found by other studies using similar data. \citet{heckman15} calculated \mout from NGC~7714 and Haro~11 assuming a constant outflow column density of 10$^{20.85}$~cm$^{-2}$, a constant outflow metallicity of 0.5~Z$_\odot$, and that all of the outflowing gas is at a constant radius that is twice the size of the star-forming region. The mass-loading factors differ by 0.02 and 0.46~dex between the two studies, within the errors of the measurement. The outflow assumptions do not heavily impact the derived mass-loading factors determined for these galaxies as compared to other studies. }

 {The fact that our empirical metal-loading factors agree with the analytic chemical evolution models of \citet{peeples11} supports the latter's conclusion that efficient outflows shape the MZR.   However, one caveat is that these models assume that accreting gas has zero metallicity.   Accretion is clearly an important process; 1~Zw~18 and Haro~11 lie below the MZR and their metallicities are often explained by recently accreted low-metallicity gas \citep{vanzee, ekta}.  However, recent simulations show that most of the gas accreted onto galaxies at z~=~0 is not pristine gas, rather it is substantially metal-enriched, recycled galactic outflows \citep{angles}.  If true, the metal-loading factors would have to be higher to match the observed MZR. More observational constraints on the accretion process are required to clarify the situation.}

 {We used the SFRs measured within the COS aperture as opposed to the global SFRs because we find that the bulk of the gas in the outflow is $< 100$~pc above the star-forming region (\autoref{tab:prof}), and thus more likely to be associated with local driving mechanisms rather than global ones.  If we replaced the SFRs with global ones, the metal-loading factors would decrease by 0.27~dex on average. SFRs calculated from a combination of the UV and IR luminosities are more robust than SFRs calculated form the UV alone \citep{hao}.  However, we do not have high-spatial resolution IR emission maps. Consequently, we assumed that the fraction of dust obscured star formation within the COS aperture is similar to the global average.  However, since we preferentially target UV-bright regions which may have lower dust attenuation than the average, this may overestimate the SFR within the COS aperture, and thereby underestimate the mass and metal-loading factors. This would impact the two most massive galaxies in our sample the most (NGC 7714, NGC 6090) which have the highest dust attenuation (\autoref{tab:sample}).}    

 {Both the UV and IR trace star formation on timescales of $\sim$100~Myr \citep{kennicutt2012}.  Analysis of the resolved stellar populations in I~Zw~18 and SBS~1415+437 suggest that a large majority of the star formation has occurred over an extended period in the past 100~Myr \citep{mcquinn, annibali}, with up to a factor of 5 variation in amplitude during this time. This 100~Myr timescale is consistent with the timescale over which energy and momentum are injected into the gas by supernovae and stellar winds \citep{claus99}.  Thus, our SFRs are appropriate to use in computing the mass and metal-loading factors of the outflow.   We have used nebular oxygen abundance measurements from the star-forming regions closest to our COS aperture to define \zism.  Thus our \zop/\zism\ values should not have any spatial or temporal biases.}

\subsection{Future work and outstanding questions}
\label{further}
Above we presented results suggesting that the metal-loading factors of nearby galactic outflows are consistent with analytic requirements to shape the MZR (\autoref{fig:zout}). A large hurdle of this study is having a statistically relevant sample of star-forming galaxies covering the full range of galaxy properties. With only seven local galaxies, the small sample size cannot definitively discern whether galaxies have metal enriched galactic outflows or if galactic outflows shape the MZR. The small size is largely because it is observational challenging to observe the four weak ISM absorption lines due to contamination by Milky Way absorption and geocoronal emission lines at low redshifts. 

Crucially, the two low-mass galaxies in the sample have contamination from Milky Way absorption, geocoronal emission, and zero-velocity absorption (see \autoref{prev}). These low-mass galaxies also have the largest parameter uncertainties due to their weak and narrow line profiles (\autoref{fig:s42}). More high-quality observations of outflows from low-mass galaxies will determine the scaling of the metal-loading factor. This will provide new constraints for galaxy simulations by developing a new empirically motivated feedback constraint (see \autoref{highz}).

While local galaxies are observable with COS, half of the total stars formed--and therefore metals produced--happened near $z \sim 2$. Therefore, these redshifts are crucial to understand why the shape of the MZR does not evolve with time, but the normalization does vary with redshift \citep{zahid}. Fortunately, at $z \sim 2$ the lines used in this analysis are redshifted into the optical, enabling a similar study of metal-loading factors in moderately redshifted galaxies.

The advent of integral field spectroscopy (IFS) emphasizes that the metallicity depends both on the mass of the galaxy and the location within the galaxy. Metallicity gradients have been discovered in a variety of galaxy types in the local universe \citep{sanchez, belfiore}. Understanding how the outflow metallicity changes with spatial location within a galaxy will demonstrate the role outflows play in creating metallicity gradients. New, high through-put, IFS instruments, like MUSE and the Keck Cosmic Web Imager (KCWI), allow for similar measurements as we presented here to be made of spatially resolved gravitationally lensed galaxies at $z > 2$. Early spatially resolved galactic outflow studies are inconclusive whether outflows vary spatially \citep{bordoloi} or remain relatively constant within a galaxy \citep{james18}. Further results will determine whether spatially varying outflows produce metallicity gradients.

Finally, the hot outflow phase is the least observed and most poorly constrained phase. The hot outflow phase is poorly constrained because current X-ray telescopes can only feasibly measure the outflow metallicities from a small sample of the nearest galaxies \citep{strickland2000, martin02, strickland09}. Future X-ray telescopes will have the collecting area and spectral resolution required to measure the metal outflows of a representative sample to determine the metallicity variation of the hot outflows. 

\section{Summary}

We conclude a series of papers studying the galactic outflow properties of seven local star-forming galaxies using restframe ultraviolet interstellar metal absorption lines. These seven galaxies span a large range in stellar mass (\mstarp) and star formation rate (SFR). In this paper, we focused on the results of {\small CLOUDY} photoionization modeling of the galactic outflows using four weak blueshifted absorption lines and the observed stellar continua as the ionizing source to determine the ionization structure and metallicity of  {the warm-phase of} galactic outflows (\autoref{cloudy}). 

We studied how the outflow metallicity (\zop) and the metal outflow rate (\mzp) scales with \mstarp. The outflow metallicity has a median \zop~=~$1.0 \pm 0.6$~Z$_\odot$ and does not vary with \mstar (left panel of \autoref{fig:z}). Outflows are metal-enriched. Galaxies with log(\mstarp)~>~9 have outflow metallicities 2.6 times larger than their ISM metallicities. Similarly, outflow metallicities from low-mass galaxies are 10-50 times larger than their ISM metallicities (left panel of \autoref{fig:zout}). Therefore, outflows remove substantially more metals from low-mass galaxies. The metal outflow rate increases with increasing \mstar as $M_\ast^{0.2\pm0.1}$ (right panel of \autoref{fig:z}). The metal-loading factor relates the rate that outflows remove metals from galaxies to the rate that star formation retains metals ($\zeta = {\rm Z}_\text{o}\times\dot{M}_\text{o}/$\zism$\times$SFR). The metal-loading factor strongly anti-correlates with \mstar (3$\sigma$; right panel of \autoref{fig:zout}), implying that outflows from low-mass galaxies remove 100 times more metals than their star formation retains. 

We then compared the observed metal-loading factors to analytic estimates of the metal-loading factors required to shape observed mass-metallicity relationships (see the red and blue curves in the right panel of \autoref{fig:zout}). The metal-loading values are qualitatively and quantitatively consistent with analytic predictions. We fit the observed scaling of the metal-loading factor with stellar mass (and circular velocity) with a power-law (\autoref{eq:zeta}) and the fitted values agree with theoretical values required to produce observed mass-metallicity relations.  This suggests that galactic outflows may shape the mass-metallicity relationship. 

The observed outflows are mostly entrained ISM (see \autoref{fig:fentrain});  even the most enriched outflows are still 78\% entrained ISM. Furthermore, since the outflows are largely entrained ISM and \zo does not vary with \mstarp, the mass-metallicity relation is largely shaped by star formation driven galactic outflows becoming less efficient at larger stellar masses.  We explored the origins of a \mstar independent outflow metallicity and tested the impact of metal-enriched outflows on the normalization of the mass-metallicity relation (see \autoref{drive}).

We also discussed some limitations of our current sample, particularly that the small sample size makes it challenging to definitively determine the metal outflow scaling relations (\autoref{prev}). Future observations of low-mass galaxies will test the driver of the stellar mass-metallicity relationship, and whether outflows have larger metallicities than their ISM (\autoref{further}). 

These observations are the first direct measurements of the outflow metallicity of the warm galactic outflow phase from a diverse sample of local galaxies. These observations emphasize that outflows shape the stellar mass-metallicity relation. The metal-loading factor provides a new observational constraint for stellar feedback prescriptions and may lead to more realistic implementations of metal-enriched galactic outflows in galaxy formation and evolution simulations. 

\section*{acknowledgements}
We thank the anonymous referee for a thoughtful and helpful report that substantially improved the quality of the manuscript. We also thank Jason X. Prochaska for helpful comments on the manuscript.
\bsp	
\label{lastpage}

\appendix
\section{EXPLORING THE PHOTOIONIZATION MODELS}
\label{cloudy_obs}
In this Appendix we explore the photoionization models. Specifically, we study how observations (dust attenuation and \rft~$=\log(N_\mathrm{\siivp}/N_\mathrm{\siiip})$) correspond to the modeled parameters (\zo and $\chi_\mathrm{\siivp}$).  In the following discussion we include the $z \sim 2.9$ gravitationally lensed galaxy from the \megasaura\ sample \citep{rigby} where the ionization structure was analyzed in a similar way as in \autoref{cloudy} \citep{chisholm18}. \citet{chisholm18} find log(U)~$=-2.01\pm 0.02$, \zo~$= 1.56\pm 0.15$~Z$_\odot$ (in the solar metallicity convention used here), \zs~$=0.2$~Z$_\odot$, \rft~$=-0.46$, $N_\text{H}=5.6\times10^{20}$~cm$^{2}$, and $\chi_\mathrm{\siivp} = 0.14$. However, at these redshifts accurate ISM metallicities have not been observed. Consequently this galaxy is not included in the mass-metallicity analysis above. In this section we consider only the photoionization relations and include this galaxy as a light-blue point in all of the plots of this section.

\subsection{The impact of dust attenuation on modeled parameters}
The {\small CLOUDY} modeled total hydrogen column density ($N_\mathrm{H}$) times \zo weakly scales with the gas-phase extinction (\ebv) measured during the stellar continuum fitting (left panel of \autoref{fig:u}). $N_\mathrm{H}\times\mathrm{Z}_\mathrm{o}$ increases with increasing \ebv\ (the significance is less than 2$\sigma$), implying that larger metal column densities lead to more extinction.

A relationship between $\mathrm{N}_\mathrm{H}\times$\zo and \ebv\ is expected if the dust-to-gas ratio depends linearly on the metallicity of the gas attenuating the stellar continuum. A similar relationship has been observed between molecular gas column density and metallicity \citep{issa, leroy11}. Using the starburst version of equation 4 from \citet{heckman11} and equation 10 from \citet{calzetti}, the expected relationship scales as
\begin{equation}
\mathrm{N}_\mathrm{H} \mathrm{Z}_\mathrm{o} =5.5\times 10^{21}~\mathrm{cm}^{-2} \mathrm{E}_\mathrm{S}\mathrm{(B-V)} .
\label{eq:ebv}
\end{equation}
This relationship is overplotted in the left panel of \autoref{fig:u}. This relationship depends on the assumed reddening law, and the normalization constant in \autoref{eq:ebv} would decrease to $3.3\times10^{21}$~cm$^{-2}$ if a SMC law was used.  {We demonstrate this difference by over-plotting the predicted SMC relationship as a dot-dashed line in the left panel of \autoref{fig:u}. Most points line along the SMC relation, while two points are along to the Calzetti relation.} \ebv\ is not an input of the photoionization modeling, but the qualitative agreement between the modeled N$_\mathrm{H}\times\mathrm{Z}_\mathrm{o}$ and the observed \ebv\ increases the confidence in the estimated ionization structure, total column density, and \zo of the photoionization modeling.

\subsection{The impact of \rft\ on modeled parameters}

The ionization fraction ($\chi_{i}$ where $i$ is a particular ion) is the fraction of the total element in a particular ionization state. The middle panel of \autoref{fig:u} shows the relationship between the modeled $\chi_\mathrm{\siivp}$ and the observed \siivp-to-\siii column density ratio (\rft$= \log(N_\mathrm{\siivp}/N_\mathrm{\siiip})$). $\chi_\mathrm{\siivp}$ scales with \rft, such that outflows with lower $\chi_\mathrm{\siivp}$ have less highly ionized silicon than partially neutral silicon (\rft$~< 0$). Conversely, outflows with larger $\chi_\mathrm{\siivp}$ have relatively more highly ionized silicon (\rft$~>0$). A $2.5\sigma$ (p-value < 0.006, R$^2$ = 0.76) relation is found between \rft\ and $\chi_\mathrm{\siiv}$ with a relation of:
\begin{equation}
\chi_\mathrm{\siiv} = (0.26\pm0.06) \mathrm{R}_{42} + \left(0.23 \pm 0.02\right) .
\label{eq:colrat}
\end{equation}
The considerable scatter in this relationship is likely because a fraction of the Si can also be found in other ionization states (mainly \siiiip) which are not accounted for in this plot. 

\begin{figure*}
\includegraphics[width = \textwidth]{cloudy_obs.pdf}
\caption{Plots of {\small CLOUDY} modeled photoionization quantities. {\it Left Panel:} The relationship between the modeled total hydrogen column density (N$_\mathrm{H}$) times the modeled outflow metallicity (\zop) and the observed gas-phase attenuation (\ebv) measured from fitting the UV stellar continuum (\autoref{cont}).  {The dashed (dot-dashed) line shows an analytic relationship between these quantities assuming a Calzetti (SMC) dust law if the dust-to-gas ratio scales linearly with \zo (\autoref{eq:ebv}).} The light-blue point is the high-redshift galaxy from the \megasaura\ sample \citep{rigby, chisholm18}. {\it Middle Panel:} The relation between the derived \siiv ionization fraction ($\chi_\mathrm{\siivp}$) and the observed \siivp-to-\siii column density (\rft~$=\log(\mathrm{N}_\mathrm{\siivp}/\mathrm{N}_\mathrm{\siiip})$). {\it Right Panel:} Relationship between the modeled $\chi_\mathrm{\siivp}$ and the modeled outflow metallicity (\zop).}
\label{fig:u}
\end{figure*}

\begin{figure*}
\includegraphics[width = \textwidth]{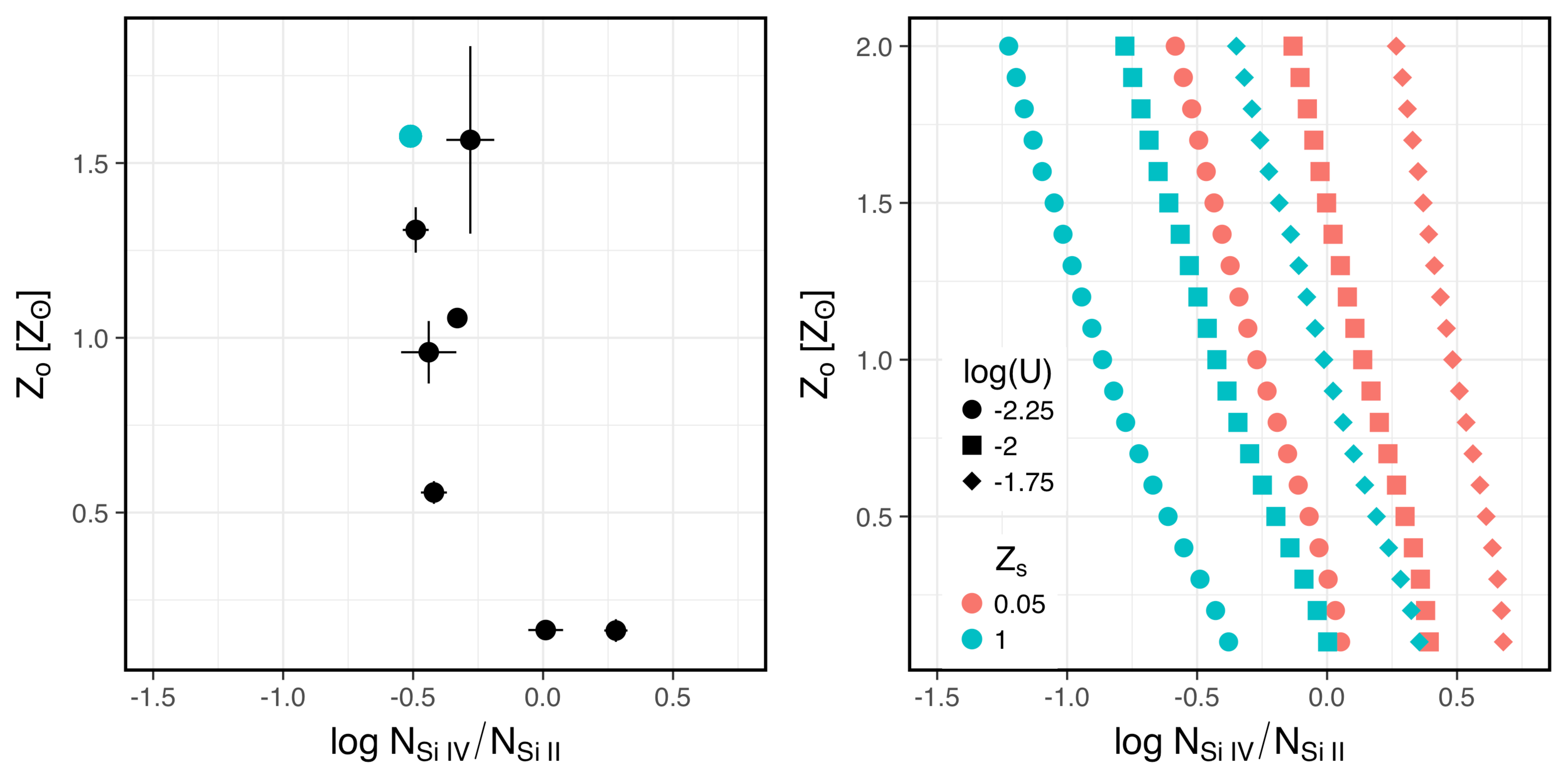}
\caption{Plots of the \siiv to \siii column density ratio (\rft~=~log($N_\mathrm{\siivp}/N_\mathrm{\siiip}$)) versus the outflow metallicity (\zop). {\it Left panel:} the relation between the observed \rft\ and the {\small CLOUDY} modeled \zop. The light-blue point is the high-redshift galaxy from the \megasaura\ sample \citep{rigby, chisholm18}. {\it Right panel:} Similar to the left panel, but for a large grid of {\small CLOUDY} models with varying ionization parameters (log(U); circles, squares and diamonds correspond to log(U) =-2.25, -2, -1.75, respectively) and stellar continuum metallicities (\zs; red and light-blue correspond to 0.05~Z$_\odot$ and 1~Z$_\odot$, respectively). Comparing the plots on the left and right relates the observed \rft\ and \zs\ values to the modeled log(U) and \zop. These plots do not illustrate the effect of stellar age.}
\label{fig:cloudy_mod}
\end{figure*}

While, \rft\ strongly correlates with $\chi_\mathrm{SiIV}$, two column densities cannot define the ionization structure and \zop. The observed \rft\ does not scale with \zo (left panel of \autoref{fig:cloudy_mod}). This is largely because \zo is not the only parameter that determines \rft, but other parameters such as U, \zs, and stellar age contribute to the ratio. We illustrate this in the right panel of \autoref{fig:cloudy_mod} by showing the large grid of {\small CLOUDY} models with varying \zop, U, and \zs. \rft\ is sensitive to U because more ionizing photons (i.e. log(U) becomes more positive) ionize the gas more (\rft\ becomes more positive). In the right panel of \autoref{fig:cloudy_mod}, the light-blue circles (log(U) = -2.25) have an \rft\ value $\sim1$~dex larger than the light-blue diamonds (log(U) = -1.75) at Z$_\mathrm{o} = 1$~Z$_\odot$.

Similarly, \rft\ is sensitive to \zs\ because lower metallicity stars have harder SEDs due to the fact that they have fewer metals in their atmospheres to absorb and scatter ionizing photons to redder wavelengths (line-blanketing). Consequently, low \zs\ stellar populations produce more ionizing photons than metal-rich stellar populations, leading to larger \rft\ values at similar \zo and log(U). This is apparent from the 0.65~dex \rft\ separation between the light-blue and red circles at Z$_\mathrm{o} = 1$ in \autoref{fig:cloudy_mod}. The age of the stellar population has a similar affect on \rft. Fortunately, the age and \zs\ in this study are constrained from the stellar population fits to the UV stellar continua (\autoref{cont}). The right panel of \autoref{fig:cloudy_mod} illustrates the complicated nature of only using two column densities to estimate the photoionization parameters, and emphasizes the need for more than two column densities to constrain U and \zop.

\subsection{The impact of modeled ionization fractions on modeled metallicities}

The final relation that we explore is between two modeled quantities: $\chi_\mathrm{\siivp}$ and \zop. Most metallicity calibrations relate the ionization structure of the gas to the metallicity because metal cooling regulates the temperature and, therefore, the ionization structure of the gas. For example, the common metallicity indicator \rtt~=~([\ion{O}{ii}]3727\AA+[\ion{O}{iii}]4959,5007\AA)/H$\beta$ \citep{pagel79}, relates the ionized oxygen emission to hydrogen recombination lines (which are emitted from the same gas as \oi emission). Similarly, the modeled $\chi_\mathrm{\siivp}$ sensitively depends on the outflow metallicity (right panel of \autoref{fig:u}), with a 3$\sigma$ relation  (p-value < 0.001; Pearson's $r$ = -0.88) between the modeled \zo and the modeled $\chi_\mathrm{SiIV}$  given as:
\begin{equation}
\chi_\mathrm{\siiv} = {0.13 \pm 0.01} \left(\frac{\mathrm{Z}_\mathrm{o}}{1~\mathrm{Z}_\odot}\right)^{-0.45\pm0.07} .
\label{eq:zfrac}
\end{equation}
This correlation illustrates the dependency of the modeled $\chi_\mathrm{\siivp}$ on \zop.

\section{Tables}
Here we give tables of the quantities used to derive the metal outflow rates of the galaxies. 

\begin{table*}
\caption{Fit properties of the \siiv line profile (see \autoref{eq:cfbeta} for details). The first column gives the name of the galaxy, the second column gives the optical depth coefficient ($\tau_0$), the third column gives the covering fraction at the initial radius ($C_f (R_\mathrm{i}$)), the fourth column gives the exponent that relates the observed velocity to the radius ($\beta$; see \autoref{eq:beta}), the fifth column gives the power-law exponent of the covering fraction ($\gamma$; \autoref{eq:cf}),  the sixth column gives the power-law exponent for the density distribution ($\alpha$; see \autoref{eq:denlaw}). These distributions are combined with the maximum observed velocity (column 7) and the photoionization models to derive the mass outflow rate (\mout; column 8) and the initial radius (\autoref{eq:tau0}; column 9).  }
\begin{tabular}{lcccccccccc}
\hline
Galaxy name & $\tau_0$ & $C_f (R_\mathrm{i})$ & $\beta$ & $\gamma$ & $\alpha$ & v$_\infty$ & \mout & $R_\mathrm{i}$ \\
 &   & & &  & & [\kmsp] & [M$_\odot$~yr$^{-1}$] & [pc] \\
  \centering(1)  & (2) & (3)& (4)& (5)& (6) & (7) & (8) & (9)  \\ 
\hline
SBS~1415+437 & $7.6 \pm 7.3$ & $0.26 \pm 0.05$ & $0.45 \pm 0.24$ & $-1.4 \pm 1.4$ & $-7.28 \pm 7.12$ & $-280$ &$0.30 \pm  0.27$ & 40 \\
1~Zw~18 & $7.1 \pm 6.8$ & $0.15 \pm 0.03$ & $0.52 \pm 0.16$ & $-2.6 \pm 2.1$ & $-10 \pm 4.5$ & $-380$ & $0.25 \pm 0.18$ & 62 \\
MRK~1486 & $4.3 \pm 1.7$ & $0.90 \pm 0.06$ & $0.44 \pm 0.11$ & $-1.3 \pm 0.42$ & $-4.4 \pm 1.5$ & $-390$ & $2.24 \pm 0.81$ & 45\\
KISSR~1578 & $3.3 \pm 0.9$ & $1.00 \pm 0.05$ & $0.50 \pm 0.10$ & $-0.88 \pm 0.21$ & $-2.8 \pm 0.5$ & $-350$ & $4.4 \pm 1.23$ & 54 \\
Haro~11 & $4.2 \pm 1.4$ & $0.95 \pm 0.03$ & $0.50 \pm 0.09$ & $-0.60 \pm 0.15$ & $-5.3 \pm 1.7$ & $-460$ & $14.63 \pm 6.24$ & 101 \\
NGC~7714 & $2.8 \pm 0.6$ & $0.97 \pm 0.02$ & $0.84 \pm 0.09$ & $-0.42 \pm 0.08$ & $-4.2 \pm 0.7$ & $-460$ & $0.71 \pm 0.25$ & 33 \\
NGC~6090 & $4.8 \pm 1.4$ & $1.00 \pm 0.04$ & $0.43 \pm 0.07$ & $-0.82 \pm 0.23$ & $-5.7 \pm 1.5$ & $-400$ & $2.96 \pm 1.00$ & 63\\
\end{tabular}
\label{tab:prof}
\end{table*}

\begin{table*}
\caption{Column densities used in the photoionization modeling. The \siiv column densities are corrected for the fitted covering fraction of the \siiv line, while the \oip, \siiip, and \sii column densities are corrected for using the \siii covering fraction. Column two gives the \oip~1302\AA\ column density, column three gives the \siiip~1304\AA\ column density, column four gives the \siip~1250\AA\ column density, and column five gives the \siivp~1402\AA\ column density. Lines that are contaminated by geocoronal emission are labelled as "Geo".  }
\begin{tabular}{lcccccccccc}
\hline
Galaxy name & log($N_\mathrm{\oip}$) & log($N_\mathrm{\siiip}$) & log($N_\mathrm{\siip}$) & log($N_\mathrm{\siivp}$) \\
 & [log(cm$^{-2}$)]   & [log(cm$^{-2}$)] & [log(cm$^{-2}$)] & [log(cm$^{-2}$)] \\
 (1)& (2)  & (3) & (4)& (5)&   \\ 
\hline
SBS~1415+437 & $15.29 \pm 0.06$ & $14.79 \pm 0.07$ & $15.61 \pm 0.06$ & $14.35 \pm 0.08$ \\
1~Zw~18 & Geo & $14.87 \pm 0.07$ & $15.34 \pm 0.05$ & $14.59 \pm 0.06$ \\
MRK~1486 & $15.29 \pm 0.03$ & $14.91 \pm 0.03$ & $15.11 \pm 0.10$ & $14.49 \pm 0.04$\\
KISSR~1578 & $14.74 \pm 0.04$ & $14.52 \pm 0.03$ & $14.65 \pm 0.40$ & $14.53 \pm 0.06$ \\
Haro~11 & $14.50 \pm 0.05$ & $14.36 \pm 0.04$ & $14.55 \pm 0.22$ & $14.64 \pm 0.02$ \\
NGC~7714 & $15.29 \pm 0.01$ & $15.06 \pm 0.01$ & $15.14 \pm 0.04$ & $14.73 \pm 0.01$ \\
NGC~6090 & $15.46 \pm 0.03$ & $15.12 \pm 0.03$ & $15.50 \pm 0.17$ & $14.63\pm0.04$  \\\
\end{tabular}
\label{tab:column}
\end{table*}

\begin{table*}
\caption{Derived photoionization properties from the {\small CLOUDY} models. Column two is the ionization parameter (U; the ratio of the photon density to the hydrogen density), column three gives the outflow metallicity (\zop), column four is the initial hydrogen number density (\no).  The next columns are quantities derived from the {\small CLOUDY} models: the neutral hydrogen column density (column 5), the total hydrogen column density (column 6), the \siiv ionization fraction (column 7), and the silicon-to-hydrogen abundance (column 8). Column 9 is the entrainment factor, or the fraction of ISM versus pure supernova ejecta within the observed outflow (defined in \autoref{eq:fe}). The metal outflow rate ($\dot{M}_\mathrm{z} = \dot{M}_\mathrm{o} \times \mathrm{Z}_\mathrm{o}$) is given in column 10.}
\begin{tabular}{lcccccccccc}
\hline
Galaxy name & log(U) & \zop & \no & log($N_{\ion{H}{i}}$) & log($N_\mathrm{H}$) & $\chi_\siiv$ & log(Si/H) & $f_e$ & \mz \\
 & [dex]   &[Z$_\odot$] &[cm$^{-3}$] & [log(cm$^{-2}$)] &[log(cm$^{-2}$)] &  &  [dex] & [\%] & [\sfrp]\\
 (1)& (2)  & (3) & (4)& (5)& (6)& (7) & (8) & (9) & (10) \\ 
\hline
SBS~1415+437 & $-2.31 \pm 0.04$ & $0.96 \pm 0.09$ & $32 \pm 10$  & 19.34 & 20.60 & 0.12 & -5.33 & 87 & $0.0042 \pm 0.0038$  \\
1~Zw~18 & $-2.22 \pm 0.06$ & $1.57 \pm 0.27$ &  $20 \pm 30$ & 19.48 & 20.63 & 0.11 & -5.11 & 78 & $0.0056 \pm 0.0041$\\
MRK~1486 & $-2.08 \pm 0.03$ & $0.56 \pm 0.03$& $30 \pm 8.9$ & 19.32 & 20.85 & 0.16 & -5.56 & 94 & $0.018 \pm 0.007$ \\
KISSR~1578 & $-1.75 \pm 0.05$ & $0.16 \pm 0.03$ & $32 \pm 5.5$ & 19.39 & 21.21 & 0.30 & -6.09 & 101 & $0.010 \pm 0.003$ \\
Haro~11 & $-1.65 \pm 0.01$ & $0.16 \pm 0.03$  & $34 \pm 6.2$ & 19.34 & 21.28 & 0.28 & -6.10 & 100 & $0.034 \pm 0.016$ \\
NGC~7714 & $-1.80 \pm 0.01$ & $1.06 \pm 0.03$  & $24 \pm 7.9$ & 18.81 & 20.97 & 0.11 & -5.28 & 89 & $0.011 \pm 0.004$\\
NGC~6090 & $-1.85 \pm 0.02$ & $1.31 \pm 0.07$ &  $19 \pm 2.4$ & 18.79 & 20.89 & 0.09 & -5.19 & 88 & $0.055 \pm 0.018$\\\
\end{tabular}
\label{tab:ion}
\end{table*}

\bibliographystyle{mnras}
\bibliography{main}

\end{document}